\documentclass[twocolumn]{aa} 
\pdfoutput=1

\usepackage{graphicx}
\usepackage{dcolumn}
\usepackage{bm}
\usepackage[colorlinks=true,linkcolor=blue]{hyperref}

\usepackage{adjustbox}
\usepackage{blindtext}
\usepackage{amsmath}
\usepackage{svg}
\usepackage[normalem]{ulem}
\newcommand{\orcid}[1]{\textsuperscript{\href{https://orcid.org/#1}{\includegraphics[width=8pt]{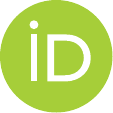}}}}

\usepackage{xcolor}
\hypersetup{
    colorlinks,
    linkcolor={red!50!black},
    citecolor={blue!50!black},
    urlcolor={blue!80!black}
}
\usepackage{booktabs}
\usepackage{natbib}
\bibpunct{(}{)}{;}{a}{}{,}
\usepackage{float}
\newcommand{\fsig}{f{\sigma_8}}

\newcommand{\qper}{q_\perp}
\newcommand{\package}[1]{{\textsc{#1}}}

\newcommand{\qpar}{q_\parallel}

\newcommand{\hmpc}{\,h^{-1}\,\text{Mpc}}
\newcommand{\ihmpc}{h\, \text{Mpc}^{-1}}

\newcommand{\rmin}{r_{\text{min}}}
\newcommand{\kmax}{k_{\text{max}}}

\newcommand{\rhMpc}{h^{-1}\,\text{Mpc}}
\newcommand{\khMpc}{h\,\text{Mpc}^{-1}}

\usepackage[switch]{lineno}

\begin{document}
\title{Growth rate measurements from a joint analysis of the large-scale galaxy clustering in Fourier and configuration space}

\author{Vincenzo Aronica\orcid{0009-0007-2476-5074}\fnmsep
      	\inst{1}\fnmsep\thanks{Corresponding author: \email{aronica@cppm.in2p3.fr}}
      	\and
      	Julian E. Bautista\orcid{0000-0002-9885-3989}\fnmsep\inst{1}
            \and 
            Arnaud de Mattia\orcid{0000-0003-0920-2947}\fnmsep\inst{2}
            \and
            Hector Gil-Marín\orcid{0000-0003-0265-6217}\fnmsep\inst{3,4,5}
      	}
   \institute{
                Aix Marseille Univ, CNRS/IN2P3, CPPM, Marseille, France
                \and 
                IRFU, CEA, Université Paris-Saclay, F-91191 Gif-sur-Yvette, France
                \and 
                Departament de Física Quàntica i Astrofísica, Universitat de Barcelona, Martí i Franquès 1, E08028 Barcelona, Spain
                \and
                Institut d'Estudis Espacials de Catalunya (IEEC), 08034 Barcelona, Spain
                \and 
                Institut de Ciències del Cosmos (ICCUB), Universitat de Barcelona (UB), c. Martí i
                Franquès, 1, 08028 Barcelona, Spain.
       	    }

   \date{}

\date{\today}

\abstract{In this work, we test a framework to perform the analysis of redshift-space distortions simultaneously in configuration and Fourier space. We test our methods with the AbacusSummit suite of N-body simulations as well as a more numerous set of approximate EZmocks, reproducing the sample of luminous red galaxies of from the Baryon Oscillation Spectroscopic Survey (BOSS) and its extension (eBOSS). 
Our clustering models are based on the effective field theory of large-scale structures in a Lagrangian frame, used in the latest results from the Dark Energy Spectroscopic Instrument. We perform a template type of analysis, including dilation parameters and the slope parameter from the ShapeFit framework.  
We find that the joint space inference yields unbiased and robust constraints on simulated datasets, consistent with results from individual spaces or previous methods to obtain consensus results.  
Our joint space analysis on the the BOSS+eBOSS LRG sample obtains \( f\sigma_8 = 0.463 \pm 0.052 \), in good agreement with the official 2020 results.
}


    \titlerunning{Joint RSD analysis in configuration and Fourier space}
    \authorrunning{V. Aronica et al. }

\maketitle

\section{Introduction}
\label{sec:intro}

Over the past decades, the study of the statistical properties of the large-scale structure (LSS) of our Universe has led to major advances in our understanding of the cosmic amounts of dark energy and dark matter as well as the initial conditions that originated such structures. Larger and denser maps of the LSS have been built thanks to the increasing power of multiplex fiber-fed spectroscopic surveys such as the cosmological programs of the Sloan Digital Sky Survey (SDSS, \citealt{eisensteinSDSSIIIMassiveSpectroscopic2011, blantonSloanDigitalSky2017}) and the Dark Energy Spectroscopic Instrument (DESI, \citealt{desicollaborationDESIExperimentPart2016, desicollaborationDESIExperimentPart2016a, desicollaborationOverviewInstrumentationDark2022}). Such surveys measure millions of high precision galaxy redshifts, spanning large fractions of our cosmic history. These maps enable estimates of the galaxy two-point functions, such as the power spectrum (in Fourier space) or the correlation function (in configuration space). The shape of such statistical measures highly depends on the expansion history of the Universe and on the growth history of initial cosmological perturbations.

The observed two-point functions of a galaxy redshift survey is anisotropic with respect to the line-of-sight (LOS) due to redshift-space distortions (RSD). Such distortions appear when converting redshifts to distances: peculiar velocities modify the observed redshifts in a coherent manner. 
On large scales, the velocity field is simply a response to the growth of density perturbations. RSD analyses aim at the measurement of such anisotropies in the two-point functions, providing constraints on the logarithmic growth rate of structure $f \equiv \mathrm{d} \ln D(a)/ \mathrm{d} \ln a$ (where $D(a)$ is the growth factor). Measurements of the growth rate of structures are a powerful cosmological probe of gravity, complementary to measurements of the expansion rate from type-Ia supernovae or baryon acoustic oscillations (BAO). 

Modelling anisotropies on large linear scales (above $\sim 80$~Mpc) is straighforward \citep{kaiser_clustering_1987}, but on intermediate to small scales (below $\sim 50$~Mpc) it can be a challenge due to non-linear clustering and a more complex galaxy-dark-matter connection. Recently, effective field theories of large-scale structures (EFTofLSS, \citealt{ivanovEffectiveFieldTheory2022}) have been able to model intermediate scale two-point clustering in redshift space. Such theories incorporate effective coefficients that depend on small-scale physics, which are then marginalised over. 

Traditionally, RSD measurements have been conducted either in configuration space 
\citep{bautistaCompletedSDSSIVExtended2021,
houCompletedSDSSIVExtended2021, 
tamoneCompletedSDSSIVExtended2020} 
or in Fourier space
\citep{gil-marin_completed_2020, de_mattia_completed_2020, neveuxCompletedSDSSIVExtended2020}.
In configuration space, the two-point correlation function $\xi(\vec{r})$ is estimated with pair-counting algorithms. In Fourier space, the power spectrum $P(\vec{k})$ is estimated by placing galaxies on a regular mesh and performing Fast Fourier transforms. 
Both approaches use the same initial dataset and, in principle, should yield the exact same cosmological results. 
In practice, however, they differ in their noise properties, on how they are impacted by systematic effects, and on the choice of scales used for inference\footnote{A specific value of $\xi(r)$ at separation \( r \) is a weighted integral of \( P(k) \) over a broad range of \( k \) values. Consequently, unless the \( r \) and \( k \) ranges used in the analyses are infinite, these approaches do not contain the same information.}. Therefore, Fourier and configuration space results are highly correlated but not identical, so there is a small gain in combining both results into a single consensus result. 

\citet{sanchez_clustering_2017} proposed a method to compute consensus results between Fourier and configuration space analyses, which works at the level of the parameter posteriors.
The method assumes that individual parameter posteriors are Gaussian, yielding also Gaussian consensus posteriors. This assumption can be broken with noisier data where posteriors in each individual analysis space are non-Gaussian. 

\citet{dumerchat_baryon_2022} developed an alternative framework to obtain consensus results, which is based on a joint fit of correlation function and power spectrum multipoles, accounting for their covariance at the inference level. Such methodology can yield non-Gaussian consensus posteriors on the fitted parameters, so it relies on less assumptions than the method by \citet{sanchez_clustering_2017}. This methodology was tested on BAO measurements of eBOSS LRG data, yielding consistent results with the official analysis.   

In this work, we extend the joint configuration and Fourier space analysis by \citet{dumerchat_baryon_2022} to an RSD analysis. 
This paper is organised as follows.  
Section~\ref{sec:dataset}
describes the real and synthetic datasets employed in our analysis, followed by a presentation of different consensus methodologies in Section~\ref{sec:methodo}. 
The theoretical model of the two-point functions used in our inferences is described in Section~\ref{sec:model}. 
In Section~\ref{sec:results_mock}, we present the results of the joint RSD analysis on mock data. 
Results on eBOSS LRG data are discussed in Section~\ref{sec:results_eboss}.

\section{Dataset}
\label{sec:dataset}

In this section, we summarize the datasets used in our RSD analyses as well as the synthetic samples used to validate our methodology. 

\subsection{The BOSS and eBOSS sample of LRGs}
\label{sec:dataset:eBOSS}

The third and fourth generations of the SDSS contained two major cosmological surveys: the Baryon Oscillation Spectroscopic Survey (BOSS, \citealt{dawsonBaryonOscillationSpectroscopic2013}) and its extension, eBOSS \citep{dawsonSDSSIVExtendedBaryon2016}, each program observed for about five years each. 
The SDSS used the 2.5-meter Sloan Foundation Telescope at Apache Point Observatory (APO; \citealt{gunn_25_2006}), which is equipped with a focal plane able to host 1000 optical fibers and two spectrographs covering the 3600 to 10400~\AA\ range in wavelengths \citep{smeeMultiobjectFiberfedSpectrographs2013}. The main goal of BOSS and eBOSS was the measurement of BAO and RSD to constrain cosmological parameters \citep{alamClusteringGalaxiesCompleted2017, alamCompletedSDSSIVExtended2021}. For that, these surveys measured millions of galaxies and quasars over a large range in redshift ($0 < z < 4$). These samples are publicly available as part of SDSS Data Release 16 \citep{ahumada16thDataRelease2020}.

To apply our framework, we consider the sample of luminous of red galaxies (LRG) from the combination of BOSS and eBOSS redshits over $0.6 < z < 1.0$. The full description of the construction of this sample for the purpose of clustering measurements can be found in \citet{rossCompletedSDSSIVExtended2020}. 
This sample is composed of 174,816 eBOSS LRG redshifts over 4,242~deg$^2$ and BOSS LRGs over 9,493~deg$^2$ over the same redshift range, yielding a total of 377,458 galaxies. Several weights are computed and assigned to each galaxy to correct for spurious density fluctuations originating from target selection, spectroscopic completeness, fiber collisions, among others. The window function of the survey is characterized by a set of unclustered points, the random catalog, containing about 20 times the number of galaxies. This sample of LRG was used to measure BAO and RSD, both in configuration and Fourier space \citep{bautistaCompletedSDSSIVExtended2021,gil-marinCompletedSDSSIVExtended2020}.

\begin{table}[t]
    \centering
    \caption{Cosmological parameters used in eBOSS and DESI mock catalogs.}
    \setlength{\tabcolsep}{4pt}
    \begin{tabular}{lcc}
        \hline
        Parameter & DESI & eBOSS \\
        \hline
        \( \Omega_m \) & 0.313 & 0.310 \\
        \( \Omega_{cdm} \) & 0.264 & 0.262 \\
        \( \omega_b \) & 0.022 & 0.022 \\
        \( h \) & 0.674 & 0.676 \\
        \( \sigma_8 \) & 0.808 & 0.808 \\
        \( n_s \) & 0.965 & 0.970 \\
        \( A_s \) & \( 2.083 \times 10^{-9} \) & $2.105 \times 10^{-9}$ \\
        \( \sum m_\nu \) [eV] & 0.059 & 0.060 \\
        \( r_{\text{drag}} \) [Mpc] & 147.09 & 147.77 \\
        \hline
    \end{tabular}
    \label{tab:cosmo_parameters_desi_eboss}
\end{table}

\subsection{Synthetic datasets}
\label{sec:dataset:mocks}

We employed two types of synthetic datasets in this work: N-body simulations and approximate methods. N-body simulations are fully non-linear and are a perfect test bench for our theoretical clustering models (section~\ref{sec:model}). Approximate mocks have less realistic clustering on small-scales but easily reproduce the survey geometry, the density of tracers and make possible the production of thousands of realisations, which are useful to test the statistical properties of our inferences and estimate sample covariance matrices. 

\paragraph{N-body simulations:} The \package{AbacusSummit} N-body simulation suite \citep{garrison_high-fidelity_2019} contains 25 independent realisations in a periodic cubic box, 
each covering a volume of $8 \, h^{-3}\mathrm{Gpc}^3$.
These mocks use the \textsc{Abacus} code \citep{garrisonABACUSCosmologicalNbody2021} to evolve $6912^3$ matter particles with mass of $2\times 10^9~h^{-1} M_\odot$ from initial conditions set at $z = 99$ from second order Lagrangian perturbation theory. Halos are identified from groups of particles using the specialised spherical-overdensity-based halo
finder CompaSO \citep{hadzhiyskaCOMPASONewHalo2022}.
We used the $z=0.8$ snapshot to match the redshift range of the LRG sample. 
The halos were populated with LRGs using a halo occupation distribution model (HOD, \citealt{yuan_desi_2023}) matching the clustering properties of DESI Early Data Release \citep{desicollaborationEarlyDataRelease2024}.
To obtain the covariance matrix of the clustering measurements of our N-body simulations, we used a set of 1000 realizations of $(2 \, h^{-1}\mathrm{Gpc})^3$ boxes run with the \package{EZmock} method \citep{chuang_ezmocks_2015}, tuned to match the clustering of the n-body simulations. Note that these EZmocks are different than those described next.

\paragraph{Approximate mocks:} We used 1000 realizations of the LRG eBOSS+BOSS survey geometry generated with the \package{EZmock} method \citep{chuang_ezmocks_2015}. This approach employs the Zel’dovich approximation to model the density field at a given redshift and populate it with galaxies. The method is computationally efficient and has been specifically calibrated to reproduce the two- and three-point statistics of the galaxy sample, including mildly non-linear scales. The angular and redshift distributions of the eBOSS LRG sample, combined with the \( z > 0.6 \) CMASS sample, were accurately replicated in these mock catalogs. A detailed description of the \package{EZmocks} LRG samples can be found in \citet{zhao_completed_2021}. These mock catalogs were crucial to the analysis, enabling the estimation of error covariance matrices for clustering measurements, the evaluation of systematic observational effects on cosmology, and the assessment of correlations between different methodologies for deriving consensus results.

In our analysis, we consider two fiducial cosmologies. Table \ref{tab:cosmo_parameters_desi_eboss} presents the fiducial cosmology used for DESI and eBOSS.

\subsection{Clustering measurements}
We used the Yamamoto estimator to estimate the power spectrum multipoles \citep{yamamoto_measurement_2006}. These are computed using \package{pypower}\footnote{\href{https://github.com/cosmodesi/pypower/tree/main}{https://github.com/cosmodesi/pypower/tree/main}}. The survey geometry is described by a window function that defines the observed regions and accounts for selection effects. We apply it by directly convolving the model power spectrum with the window function in Fourier space, following \citet{beutler_interpreting_2019,beutler_unified_2021}. The window function is computed with \package{pypower}.
The correlation function multipoles are estimated using the Landy--Szalay estimator \citep{landy_bias_1993}, and are computed using \package{pycorr}\footnote{\url{https://github.com/cosmodesi/pycorr}}.

\section{Methodology for consensus results}
\label{sec:methodo}

In this section we describe our framework to derive consensus results from a joint RSD analysis in Fourier and configuration space. We first introduce the inference algorithms. Second, we remind the basics of the Gaussian approximation method to obtain consensus results. Lastly, we describe our joint analysis in detail. 

\subsection{Parameter inference} 
\label{sec:methodo:inference}

In this work, we are ultimately interested in the inference of the growth rate of structure $f$, or its combination with the amplitude of linear perturbations 
$\sigma_8$, as well as the dilation parameters $\qper,\qpar$ describing deviations from the assumed fiducial cosmology. These parameters are defined in section~\ref{sec:model}. 

The likelihood function $\mathcal{L}$ of observing a data vector given a model obtained from a set of parameters $\theta$ can be written as:
\begin{equation}
    -2 \ln \mathcal{L}(d|\theta) = \sum_{i,j}^{N_\mathrm{bins}} \Delta_i(\theta) \hat{\Psi}_{ij} \Delta_j(\theta) + c = \chi^2 + c,
    \label{eq:likelihood}
\end{equation}
where $N_\mathrm{bins}$ is the dimension of the data vector, $\Delta_i(\theta) \equiv d_i-m_i(\theta)$ is the data-model difference in bin $i$. 
The $\hat{\Psi}$ is an estimate of the precision matrix, computed as $\hat{\Psi}=(1-D_H)\hat{C}^{-1}$, where $\hat{C}$ is an estimate of the data vector covariance matrix and $(1-D_H)$ is a correction factor, described next. 

In this work, our data vector is either the correlation function or power spectrum multipoles. In our joint analysis, the data vector is a concatenation of the multipoles from both spaces. Our covariance matrices $C$ are estimated from a set of synthetic datasets (also referred to as \emph{mocks} hereafter) reproducing our survey geometry and clustering, so accounts for both cosmic variance and shot-noise contributions. From $N_\mathrm{mock}$ mock measurements of two-point functions, the sample covariance is simply defined as
\begin{equation}
\hat{C}_{ij} = \frac{1}{N_{\text{mock}} - 1}\sum_k^{N_{\text{mock}}}
\left(d_i^k - \bar{d_i}\right)
\left(d_j^k - \bar{d_j}\right),
\label{eq:sample_covariance_matrix}
\end{equation}
where $\bar{d_i}$ is the sample average of the data vector over all mocks.
\begin{table}[t]
\caption{Correction factors used in the three analyses conducted in this study.}
    \centering
    {\footnotesize
    \begin{tabular}{c c c c c c c}
    \hline
    \hline
      Analysis  & $N_\mathrm{par}$ & $N_\mathrm{bins}$ & $N_\mathrm{mock}$ & $D_H$ & $m_1$  & $m_2$\\
    \hline
    CS  & 8 & 93 & 1000 & 0.06 & 1.05 & 1.14\\
    FS  & 13     & 108 & 1000 & 0.04 & 1.03 & 1.09\\
    JS & 13 & 201 & 1000 & 0.10 & 1.11 & 1.27 \\
    JS$_\mathrm{sep}$ & 18 & 201 & 1000 & 0.10 & 1.10 & 1.26 \\
    \end{tabular}
    }
    \label{tab:m1m2}
    \tablefoot{$D_H$ is the Hartlap correction factor for an unbiased precision matrix, $m_1$ adjusts the estimated covariance matrix of the parameters, while $m_2$ scales the variance of the best-fit parameters across a set of mocks. $N_\mathrm{par}$ is the total number of parameters fitted, $N_\mathrm{bins}$ is the total size of the data vector and $N_\mathrm{mock}$ is the number
of mocks used in the estimation of the covariance matrix.}
\end{table}
Since we use a finite number of mocks, the correction factor takes the form $D_H \equiv (N_\mathrm{bins} + 1)/(N_{\text{mock}} - 1)$ to correct for the biased estimate of the precision matrix \citep{hartlapWhyYourModel2007}. Also, each element of $\hat{C}$ is affected by its own uncertainty
which shall be appropriately propagated to the parameter uncertainties. 
Following \citet{percivalClusteringGalaxiesSDSSIII2014}, this correction is done by scaling factors given by:
\begin{eqnarray}
    m_1 &=& \frac{1 + B(N_{\text{bins}} - N_{\text{par}})}{1 + A + B(N_{\text{par}} + 1)} \\
    m_2 &=& \frac{m_1}{1 - D_H} 
\end{eqnarray}
where \( D_H \) is the Hartlap factor defined above, and
\begin{eqnarray}
    A \equiv \frac{2}{(N_{\text{mock}} - N_{\text{bins}} - 1)(N_{\text{mock}} - N_{\text{bins}} - 4)} \\
    B \equiv \frac{(N_{\text{mock}} - N_{\text{bins}} - 2)}{(N_{\text{mock}} - N_{\text{bins}} - 1)(N_{\text{mock}} - N_{\text{bins}} - 4)}
\end{eqnarray}
The factor $m_1$ is applied directly to the estimated covariance matrix of the parameters for a specific measurement. Conversely, the factor $m_2$ is used to scale the standard deviation of a parameter across a series of mocks. The values of the parameters $m_1$ and $m_2$ of our baseline analysis are given in Table \ref{tab:m1m2} for configuration space (CS), Fourier space (FS) and joint space (JS).

We assumed the covariance matrix to be independent of model parameters and it remains fixed during the inference. Therefore, the $c$ term in Eq.~\ref{eq:likelihood} is simply an additive constant. 
We use two approaches for parameter estimation: a frequentist and a Bayesian one. 
    \paragraph{Frequentist Approach:} 
    Best-fit parameters are determined by minimizing $\chi^2$ using a quasi-Newton minimizer implemented in \package{iMinuit}\footnote{\href{https://iminuit.readthedocs.io/}{https://iminuit.readthedocs.io/}}. Parameter uncertainties are estimated by identifying intervals where $\chi^2$ increases by one unit, corresponding to a $68\%$ confidence level. 
    This approach does not assume Gaussianity of the parameter posterior distribution, so uncertainties can be asymmetric.
    \paragraph{Bayesian Approach:} 
    The posterior probability distribution of the parameters $\theta$ given the observed data \(d\) is given by the Bayes' theorem, $P(\theta | d) = {\mathcal{L}(d|\theta) P(\theta)}/P(d)$,
    where $\mathcal{L}$ is the likelihood from Eq.~\ref{eq:likelihood}, \(P(\theta)\) is the prior and $P(d)$ is the evidence. 
    The likelihood is explored using a Markov Chain Monte Carlo (MCMC) algorithm, implemented in  \package{emcee}\footnote{\href{https://emcee.readthedocs.io/}{https://emcee.readthedocs.io/}} \citep{foreman-mackeyEmceeMCMCHammer2013}. We run chains in parallel and assess their convergence using the Gelman-Rubin criterion \citep{gelmanInferenceIterativeSimulation1992}.

For most test cases, we checked that both approaches give consistent results. The first approach is faster and used to quickly obtain constraints of a small subset of parameters, which can be run on large samples of mocks, while the second one is slower but accurately describes the correlated posterior distributions of all paremeters. 

We used the \package{desilike}\footnote{\href{https://github.com/cosmodesi/desilike/tree/main}{https://github.com/cosmodesi/desilike/tree/main}} framework, a public tool developed by the DESI collaboration that integrates theory modeling, observable definitions, likelihood computation, theory emulation, and posterior sampling and profiling. These tools were employed in the main cosmological results from DESI 
\citep{desi_collaboration_desi_2025-1,
desi_collaboration_desi_2024-3,
desicollaborationDESI2024IV2025,
desicollaborationDESI2024FullShape2024,
desicollaborationDESI2024VI2024,
desicollaborationDESI2024VII2024}.

\subsection{Consensus results from the Gaussian approximation method}
\label{sec:methodo:gaussian_approx}

\citet{sanchez_clustering_2017} proposed a method to obtain consensus results from configuration-space (CS) and Fourier-space (FS) analyses. This method combines posteriors distributions of the parameters of interest, assuming they are Gaussian and also yielding Gaussian consensus porsteriors. We refer to this method as the Gaussian Approximation (GA) hereafter. 

In the case of RSD analysis, the parameters of interest are the normalised growth rate of structures $\fsig$ and the dilation parameters $\qper,\qpar$. Each analysis produces its own posterior which is approximated by a three-dimensional Gaussian centered in $\theta = [\fsig, \qpar, \qper]$ with a $3\times 3$ parameter covariance $C_\theta$. The consensus result is obtained by constructing a $6\times 6$ parameter covariance, where the off-diagonal blocks (i.e., the cross-covariance between parameters in CS and FS) are estimated from mock catalogs. The consensus result is given by $\theta^\mathrm{GA}$ and its $3\times3$ covariance $C^\mathrm{GA}_\theta$. 

Furthermore, \citet{bautistaCompletedSDSSIVExtended2021} argued that the cross-covariance block obtained from a set of mocks is representative of the ensemble, but not of a particular realisation. An adjustement of these off-diagonal blocks was proposed, based on the uncertainties of the data realisation itself. This technique was used to obtain consensus between configuration and Fourier space analyses as well as between BAO and RSD constraints (each yielding dilation parameters). A full description of the method we use in this work can be found in \citet{bautistaCompletedSDSSIVExtended2021}.

\subsection{Consensus results from a joint space analysis}
\label{sec:methodo:joint_space}

An alternative to the Gaussian approximation for obtaining consensus results is to perform a joint fit of both configuration-space and Fourier-space two-point functions, using a single set of parameters. The cross-covariance between multipoles in CS and FS is estimated from mock catalogs. 
We refer to this approach as the Joint space (JS), hereafter. \citet{dumerchat_baryon_2022} investigated the performance of JS fits in the context of BAO analyses. In this work, we extend that methodology to RSD analyses.

In addition to the cosmological parameters of interest ($\fsig,\qper,\qpar$), our models contain several physical parameters which are simultaneously adjusted and marginalised over. Ideally, these nuisance parameters should be consistent across CS and FS analyses. In practice, we see discrepancies on the constraints of such parameters when analysing mock catalogs in either CS or FS. These discrepancies motivated us to create versions of the JS analysis where some of the nuisance parameters are not shared, i.e., each space contains separate versions of the same nuisance parameter. We refer to this model as JS$_\mathrm{sep}$ or simply JS, as it will be our baseline choice. 

Table~\ref{tab:js} summarizes the variations considered in our JS analysis. The parameters are described in section~\ref{sec:model}. 

\begin{table}[t]
\caption{Variations of the joint-space analyses. }
    \centering
    \begin{tabular}{c c c c }
    \hline
      Name & Common  & Separate & $N_\mathrm{par}$ \\
    \hline
    \hline
    $\mathrm{JS}_\mathrm{sep}$ & $\qpar, \qper, f, m, b_1$ & $ b_2,b_s, \alpha_i$  & 18  \\
    JS & ${\qpar, \qper, f,m ,b_1, b_2,b_s,\alpha_i}$ & -  &  13\\
    \hline
    \end{tabular}
\label{tab:js}
\tablefoot{The second and third columns
indicate which parameters of the model are common and separate between Fourier and
configuration spaces. $N_\mathrm{par}$ column indicates the total number of free parameters in the fits (taking into account the 3 stochastic parameters that are only present in Fourier space).}
\end{table}

\section{Clustering model}
\label{sec:model}

We use a Lagrangian perturbation theory (LPT) model to describe the observed clustering in both configuration and Fourier space. This model is based on the Lagrangian formalism developed in \citet{chen_consistent_2020,chen_redshift-space_2021}. A more detailed presentation is provided in Appendix \ref{sec:model_app}.

\subsection{EFTofLSS and LPT in redshift space}
\label{sec:model:lpt}

In the Lagrangian picture, cosmological structure formation is modeled by tracking the trajectories of fluid elements, rather than evolving the density and velocity fields directly. The key quantity in LPT is the displacement field \(\bm{\Psi}(\bm{q}, \eta)\), which maps the initial Lagrangian coordinates \(\bm{q}\) (at some early conformal time \(\eta_0\)) to Eulerian positions \(\mathbf{x}(\bm{q})\) at time \(\eta\):  
\begin{equation}\label{eq_sec_2:displament_def}
\mathbf{x}(\bm{q}, \eta) = \bm{q} + \bm{\Psi}(\bm{q}, \eta),
\end{equation}
with the initial condition \(\bm{\Psi}(\bm{q}, \eta_0) = 0\). The field \(\bm{\Psi}(\bm{q}, \eta)\) fully captures the evolution of the fluid element labeled by \(\bm{q}\). At first order, the displacement is given by the Zel’dovich approximation \citep{zeldovich_gravitational_1970}:
\begin{equation}\label{eq_sec_2:Zeldovich}
    \bm{\nabla}_{\bm q} \bm\Psi^{(1)} = - D_{+}(\eta)\delta_0(\bm q),
\end{equation}
or equivalently,
\begin{equation}\label{eq_sec_2:Zeldovich_2}
\bm{\Psi}(\bm{q}, \eta) = iD_+(\eta) \int \frac{d^3k}{(2\pi)^3} \frac{\bm{k}}{k^2} \delta_0(\bm{k}) e^{i\bm{k} \cdot \bm{q}},
\end{equation}
where \(D_{+}(\eta)\) is the linear growth factor and \(\delta_0(\bm{q})\) the initial density contrast.

In this work, we use the Effective Field Theory of Large-Scale Structure (EFTofLSS) in its Lagrangian formulation. This involves smoothing the displacement field and systematically incorporating small-scale physics through an effective expansion (see Equation \ref{eq_sec_2:evo_psi_app} in Appendix).

This leads to the Zel’dovich power spectrum:
\begin{equation}
\label{eq_sec_2:pk_ZA}
    P_{\rm Zel} (\bm k)= \int d^3\bm q\, e^{i\bm k \cdot \bm q} \left[e^{-k_i k_j [\Psi_{ij}(\bm{0}) - \Psi_{ij}(\bm{q})]} - 1\right],
\end{equation}
where \(\Psi_{ij}(\bm{q})\) is related to the linear power spectrum via:
\begin{equation}
  \Psi_{ij}(\bm{q}) = \frac{1}{(2\pi)^3} \int d^3k \, \frac{k_i k_j}{k^4} P_{\rm lin}(k) e^{-i\bm{k} \cdot \bm{q}}.
\end{equation}

Equation \eqref{eq_sec_2:pk_ZA} describes the nonlinear mapping of the initial linear power spectrum into the evolved power spectrum via LPT.

Redshift-space distortions (RSD) are incorporated in the Lagrangian formalism by modifying displacements along the line-of-sight (LOS) direction \(\hat{n}\), accounting for galaxy peculiar velocities:
\begin{equation}
    \bm\Psi \to \bm\Psi^s = \bm\Psi + \frac{\hat{n}(\bm v \cdot \hat{n})}{\mathcal{H}},
\end{equation}
where \(\bm v\) is the peculiar velocity and \(\bm\Psi^s\) denotes the redshift-space displacement vector.

Finally, the density of biased tracers is modeled using a Lagrangian bias prescription, where the tracer density is given by a bias functional of the initial fields. This treatment is detailed in Appendix \ref{sec:model:rsd_app}.

The final expression of the theoretical model for the redshift-space galaxy power spectrum, accounting for bias in LPT and EFTofLSS, can be found in \citet{chen_redshift-space_2021,vlah_gaussian_2016,carlson_convolution_2013}. In a compact form it writtes:
\begin{equation}
\label{eq:model_pk_velo}
\begin{aligned}
     P_s(\bm k, \mu) &= P_{s,zel} (\bm k) + P_{s,1-loop}(\bm k)\\
    & + k^2(\alpha_0 + \alpha_2\mu^2 + \alpha_4\mu^4)P_{s,zel} (\bm k) \\
    & + R_h^3 (1 + \sigma_2k^2\mu^2 + \sigma_4 k^4\mu^4)   
\end{aligned}
\end{equation}
where \(P_{s,zel} (\bm k)\) is the linear in LPT (Zeldovich) power spectrum, \(P_{s,1-loop}(\bm k)\) is the 1-loop LPT power spectrum, \(\alpha_i\) are the counterterms, and \(\sigma_i\) represent stochastic contributions, which scale with the typical size of halo/galaxy formation \(R_h\). The parameter \(\mu = \hat{k} \cdot \hat{n}\) represents the cosine of the angle between the wavevector \(\bm{k}\) and the line of sight. In what follows, the stochastic terms will be simplified using \(sn_i\) notation, where \(sn_0 = R_h^3\), \(sn_2 = R_h^3\sigma_2\), and \(sn_4 = R_h^3\sigma_4\). Counterterms and stochastic contributions include components such as the shot noise and Finger-of-God (FoG) effects.

In our work, the model is computed using the \package{velocileptors}\footnote{\href{https://github.com/sfschen/velocileptors}{https://github.com/sfschen/velocileptors}} package, integrated within \package{desilike}\footnote{\href{https://github.com/cosmodesi/desilike/tree/main}{https://github.com/cosmodesi/desilike/tree/main}}. The Fourier-based \package{velocileptors} code  implements the redshift-space power spectrum at one-loop order in Lagrangian Perturbation Theory (LPT) with resummation up to a scale \(k_\mathrm{IR}\) \citep{chen_redshift-space_2021}. In addition to the three stochastic parameters and three counterterm parameters, the bias model includes three parameters: \(b_1\), \(b_2\), and \(b_s\). We decided to fix $b_3$ to zero as it is expected to be small in our analysis and is quite degenerate with the counterterms \citep{desicollaborationDESI2024VI2024,maus_analysis_2024}. 

When performing cosmological inference each likelihood evaluation requires calculating loop-correction terms, which involve computationally expensive 2D integrals ($\sim$ 1s per integral). Since these calculations are repeated tens of thousands of times, a fourth-order Taylor expansion emulator is employed to accelerate the process. This emulator interpolates pre-evaluated grid models, avoiding repeated loop integral evaluations and significantly reducing computation time. More information about the \package{velocileptors} model can be found in \citet{desicollaborationDESI2024VI2024}. We also speed up our MCMC fits by analytically marginalizing over the linear nuisance parameters in our model, i.e. the parameters of the stochastic and counterterm.

The choice of priors, particularly for nuisance parameters, can influence cosmological posteriors through two main effects: the {prior weight effect} (PWE), where the prior pulls the posterior away from the data, and the {prior volume effect} (PVE), where marginalization over many nuisance parameters shifts the posterior due to large-volume regions in parameter space.
To mitigate these effects, we follow the conservative strategy of \citet{desi_collaboration_desi_2024-1, maus_analysis_2024}, applying Gaussian priors on nuisance parameters. We reparameterize galaxy bias in terms of \( (1 + b_1)\sigma_8(z),\, b_2\sigma_8^2,\, b_s\sigma_8^2 \), reducing projection effects and improving robustness at low \(\sigma_8\).
For stochastic terms (\( \text{sn}_{i=0,2,4} \)) and counterterms (\( \alpha_{i=0,2,4} \)), we use Gaussian priors motivated by physical arguments. Shot-noise terms are scaled from Poisson expectations and effective velocity dispersion, while counterterms are centered at zero with width \(12.5\,h^{-2}\,\mathrm{Mpc}^2\), corresponding to 50\% of the correction at \( k_{\text{max}} = 0.20\,h\,\mathrm{Mpc}^{-1} \). The priors are given in Table \ref{tab:prior}.

\begin{table}[t]
\centering
\caption{Priors on the cosmological and non-cosmological parameters used in our analysis.}
{\footnotesize
\begin{tabular}{ll}
\hline
\hline
\multicolumn{2}{c}{{Cosmological parameters}} \\
\midrule
$\qpar$ & $\mathcal{U}[0.8, 1.2]$ \\
$\qper$  & $\mathcal{U}[0.8, 1.2]$ \\
$f/f_{\text{fid}}$    & $\mathcal{U}[0.0, 2.0]$ \\
$dm$                   & $\mathcal{U}[-3, 3]$ \\
 \\
\hline
\hline
\multicolumn{2}{c}{{Non-cosmological parameters}} \\
\hline
$(1 + b_1)\sigma_8$   & $\mathcal{U}[0,3]$ \\
$b_2\sigma_8^2$       & $\mathcal{N}[0, 5^2]$ \\
$b_s\sigma_8^2$       & $\mathcal{N}[0, 5^2]$ \\
$\alpha_0$            & $\mathcal{N}[0, 12.5^2]$ \\
$\alpha_2$            & $\mathcal{N}[0, 12.5^2]$ \\
$\alpha_4$            & $\mathcal{N}[0, 12.5^2]$ \\
$\text{sn}_0$         & $\mathcal{N}[0, 2^2] \times 1/\bar{n}_g$ \\
$\text{sn}_2$         & $\mathcal{N}[0, 5^2] \times f_{\text{sat}}\sigma_{1,\text{eff}}^2 / \bar{n}_g$ \\
$\text{sn}_4$         & $\mathcal{N}[0, 5^2] \times f_{\text{sat}}\sigma_{1,\text{eff}}^4 / \bar{n}_g$ \\

\end{tabular}
}
\label{tab:prior}
\tablefoot{Here, $\mathcal{U}$ stands for a uniform prior, and $\mathcal{N}(\mu, \sigma^2)$ for a Gaussian prior with mean $\mu$ and variance $\sigma^2$. The cosmological parameters are listed in the top sub-panel, and the non-cosmological parameters are listed in the bottom sub-panel. For the stochastic parameters, the prior width on sn$_0$ is scaled by the Poissonian shot-noise ($1/\bar{n}_g$), while the prior widths for sn$_2$ and sn$_4$ are scaled by the expected fraction of satellite galaxies, the effective velocity dispersion of the satellites, and the Poissonian shot-noise: $f_{ sat} \sigma^2_{ 1\,eff} / \bar{n}_g$, $f_{ sat} \sigma^4_{ 1\,eff} / \bar{n}_g$, respectively (see \citet{maus_analysis_2024} for details).}
\end{table}

\subsection{Shapefit}
\label{sec:model:shapefit}

In this work, we use the {Shape-Fit} approach, which builds on the standard fit method by taking a linear power spectrum template evaluated at a reference cosmology with a set of parameters. ShapeFit captures early-time physics through the slope of the power spectrum, as highlighted by \citet{brieden_shapefit_2021}. They proposed an extension to the standard method by modelling the slope of the linear power spectrum using a logarithmic rescaling, modifying the reference template as:
\begin{equation}
\ln \left( \frac{P'_{\text{ref}}(k)}{P_{\text{ref}}(k)} \right) = \frac{m}{a} \tanh \left[ a \ln \left( \frac{k}{k_p} \right) \right] + n  \ln \left( \frac{k}{k_p} \right),
\end{equation}
where \(a = 0.6\) and \(k_p = 0.03 \, h \, \text{Mpc}^{-1}\) are fixed values following \citet{brieden_shapefit_2021}. The parameters \(m\) and \(n\) correspond to the scale-dependent and scale-independent slopes, respectively. They encode the slope of the power spectrum ({"Shape"}). This ansatz was designed to replicate the effects of varying \(\omega_b\), \(\omega_m\), and \(n_s\) on the power spectrum's shape. In this work, we vary only \(m\), keeping \(n = 0\) fixed, as both show a strong anti-correlation \citep{desi_collaboration_desi_2024-1}.

In addition to the free parameters of the LPT EFT model presented in Eq.~\ref{eq:model_pk_velo}, we include \(dm = m - m_{\text{fid}}\) as an additional free parameter in our analysis

\subsection{The distance-redshift relationship}
\label{sec:model:distance_redshift}

In the context of full shape analysis, we want to model the anisotropies of the two-point measurements. To do so, we define the multipoles of the power-spectrum  $P_{\ell}(k)$ 
by integrating the model power spectrum (Eq.~\ref{eq:model_pk_velo}) over $\mu_k$ 
weighted by the Legendre polynomials $L_{\ell}(\mu_k)$:
\begin{equation}
P_{\ell}(k) = \frac{2\ell + 1}{2} \int_{-1}^{1} P(k, \mu) L_{\ell}(\mu) d\mu,
\end{equation}
The correlation function multipoles $\xi_{\ell}(r) $ are obtained by Hankel transforming the $P_{\ell}(k)$:
\begin{equation}
\xi_{\ell}(r) = \frac{i^{\ell}}{2\pi^2} \int_{0}^{\infty} k^2 j_{\ell}(kr) P_{\ell}(k) dk,
\end{equation}
where $j_{\ell}$ are the spherical Bessel functions. .

The fiducial cosmology used to convert angular positions and redshifts into distances might not match the true underlying cosmology of the observational or simulated data. Such discrepancies result in additional anisotropies often referred to as Alcock-Paczynski (AP)  distortions \citep{alcock_evolution_1979}. 
We introduce the scaling parameters $(\qper, \qpar)$, which adjust the observed radial and transverse wavenumbers $ (k_{\parallel}, k_{\perp}) $ to the true wavenumbers $ (k'_{\parallel}, k'_{\perp}) = (k_{\parallel}/\qpar, k_{\perp}/\qper) $. \cite{ballinger_measuring_1996} gave the transformation that maps  the space $ (k, \mu) = \left( \sqrt{k_{\parallel}^2 + k_{\perp}^2}, k_{\parallel}/k \right) $ to the space $ 
(k', \mu') $ :
\begin{equation}
k' = \frac{k}{\qper} \left[ 1 + \mu^2 \left( \frac{\qper^2}{\qpar^2} - 1 \right) \right]^{1/2},
\end{equation}
\begin{equation}
\mu' = \frac{\mu \qper}{\qpar} \left[ 1 + \mu^2 \left( \frac{\qper^2}{\qpar^2} - 1 \right) \right]^{-1/2}.
\end{equation}
Then we include the AP effect in the power spectrum multipoles as:
\begin{equation}
P_{\ell}(k) = \frac{2\ell + 1}{2\qpar\qper^2} \int_{-1}^{1} d\mu P_g (k'(\mu), \mu'(\mu)) \mathcal{L}_{\ell}(\mu).
\end{equation}

The scaling parameters, which might indicate a shift of the BAO feature, are defined as :  
\begin{equation}
q_{\parallel} = \frac{D_H(z_{\text{eff}})}{D_{H}^{\text{fid}}(z_{\text{eff}})},
\end{equation}
\begin{equation}
q_{\perp} = \frac{D_M(z_{\text{eff}})} {D_{M}^{\text{fid}}(z_{\text{eff}})},
\end{equation}
"fid" superscript stands for the values corresponding to fiducial cosmology, 
$D_H^{\text{fid}}(z_{\text{eff}})$ 
the Hubble distance and $D_M^{\text{fid}}(z_{\text{eff}})$ 
the comoving angular diameter distance given at the effective redshift of the galaxy sample 
$z_{\text{eff}}$.

\section{Results on mock catalogs}
\label{sec:results_mock}

In this section, we use the average of the 25 LRG cubicbox mocks from \package{AbacusSummit} N-body simulations and the 1000 eBOSS \package{EZmock} LRG realizations (described in Section \ref{sec:dataset:mocks}) to test our methodology.

\subsection{Fits to N-body cubic-box LRG mocks}
\label{sec:results_mock:nbody}

The AbacusSummit suite of N-body simulations have realistic clustering so we use them to test the performance of our models, quantifying their biases and determining a valid range of scales. The halos are populated with luminous red galaxies using an HOD model that matches the clustering measured on early DESI data.

Our data vectors are the multipoles of the correlation function and power spectrum, averaged over the 25 realisations, reducing statistical noise and providing a more robust test of the theoretical model. The covariance matrix is obtained by performing the same measurements on a set of 1000 cubic-box \package{EZmocks}, which aim at reproducing the same LRG samples of the N-body simulations. We scale this covariance matrix to match the volume corresponding to 25 times $8~(\mathrm{Mpc}/h)^3$. We used the Bayesian approach for our inferences (see section~\ref{sec:methodo}). The best-fit parameter values correspond to the median of the posterior distribution and the uncertainties are derived from the central 68\% confidence level.

We start by analysing the scales used in our fits. 
In configuration space (CS), we use separations between $r_\mathrm{min}$ and $r_\mathrm{max}$, while in Fourier space (FS), we use wavenumbers between $k_\mathrm{min}$ and $k_\mathrm{max}$. To test the performance of our non-linear modelling, we increasingly fit for smaller scales: we vary $r_\mathrm{min}$ in CS and $k_\mathrm{max}$ in FS. The large-scale limit is fixed to  $r_\mathrm{max} = 150\,h^{-1}\mathrm{Mpc}$ in CS and \(k_\mathrm{{min}} = 0.02\,h\,\mathrm{Mpc}^{-1}\) in FS. 
\begin{figure}[t]
    \centering
    \includegraphics[clip, trim=0.3cm 0.3cm 0.2cm 0.1cm,width=\columnwidth]{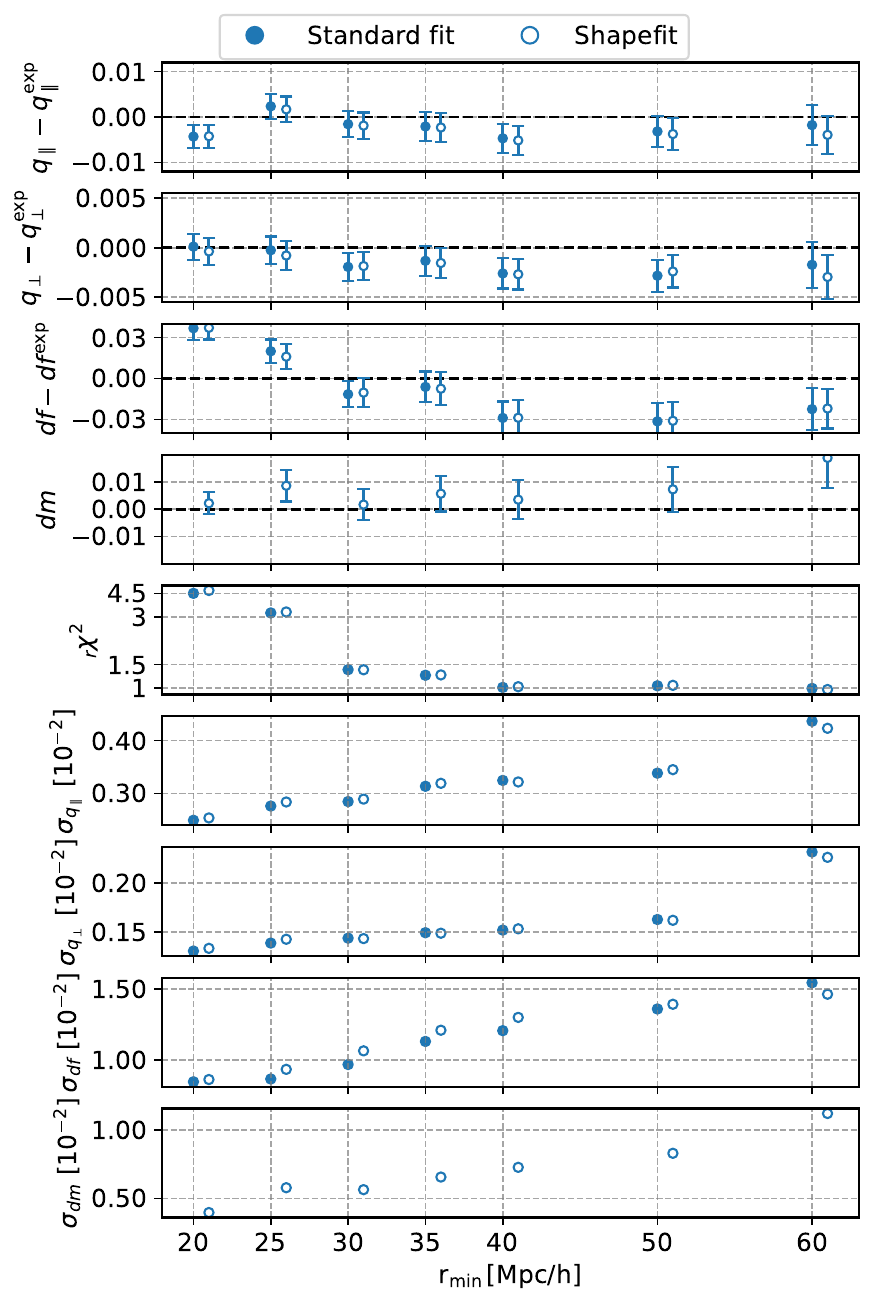}
    \caption{Configuration space RSD results from fits to the average clustering of 25 LRG cubic-box n-body simulations as a function of the minimum separation scale, $r_\mathrm{{min}}$. 
    Best-fit values of $\qpar$, $\qper$, $df$ and $dm$ are compared to their expected values in the top panels. The mid panel shows the reduced chi-squared ${}_r\chi^2$, and the bottom three panels display estimated uncertainties $\sigma_{\qpar}$, $\sigma_{\qper}$, $\sigma_{d f}$ and $\sigma_{d m}$. 
    Filled markers represent the baseline fit, while empty markers indicate fits that include the additional ShapeFit parameter $m$.}
    \label{fig:scale_rmin}
\end{figure}

\begin{figure}[t]
    \centering    
    \includegraphics[clip, trim=0.3cm 0.3cm 0.2cm 0.1cm,width=\columnwidth]{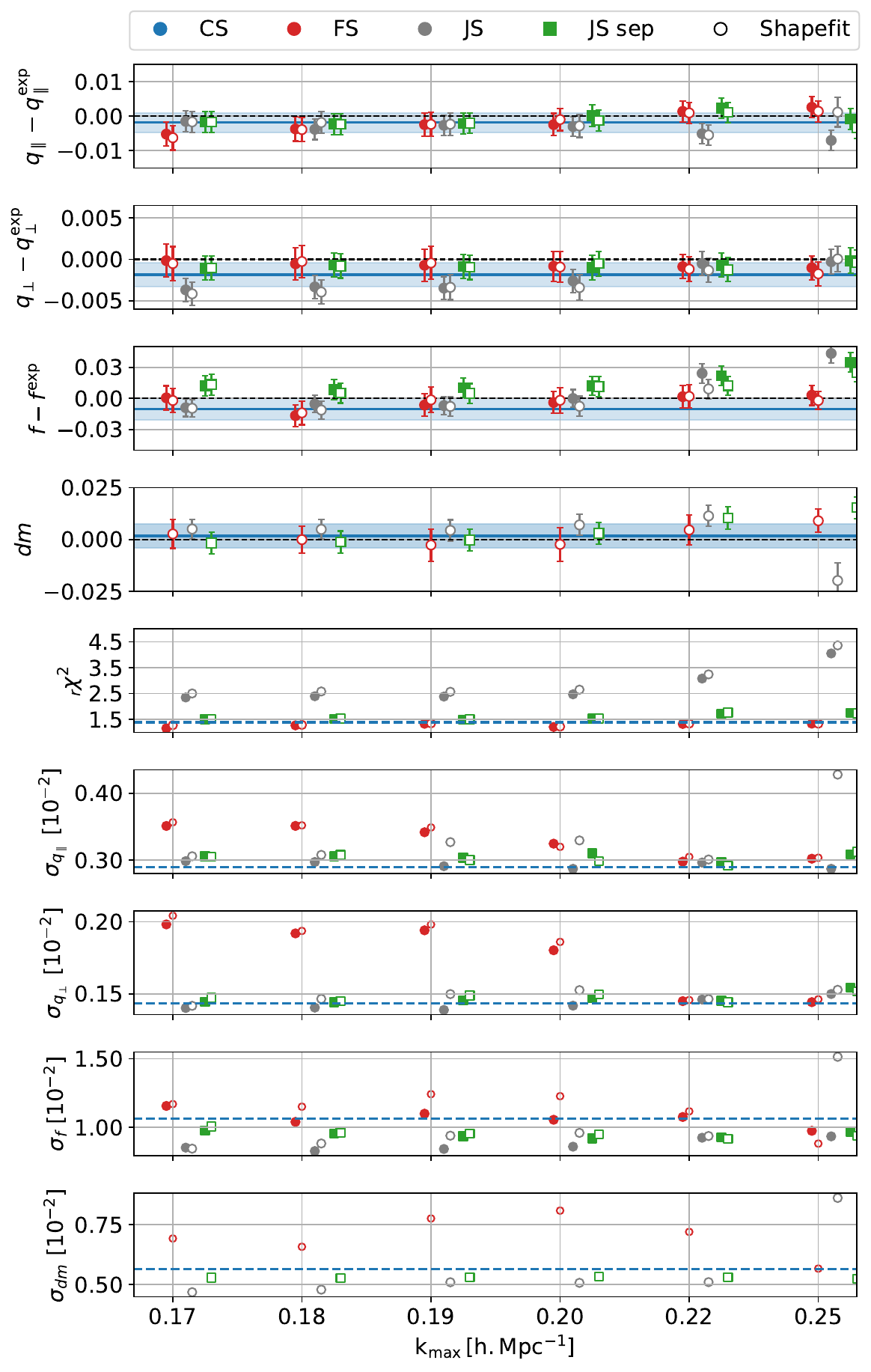}
\caption{Same as Figure~\ref{fig:scale_rmin} but for Fourier space (FS, red circles) and the joint space analyses (JS and JS$_\mathrm{sep}$, grey circles and green squares respectively). 
For reference, CS results for $r_\mathrm{{min}} = 30\rhMpc$ are shown in blue.}
    \label{fig:scale_kmax_total}
\end{figure}

\begin{figure*}[t]
    \centering
    \includegraphics[clip,trim=2cm 1cm 2.5cm 2.5cm, width=0.8\textwidth]{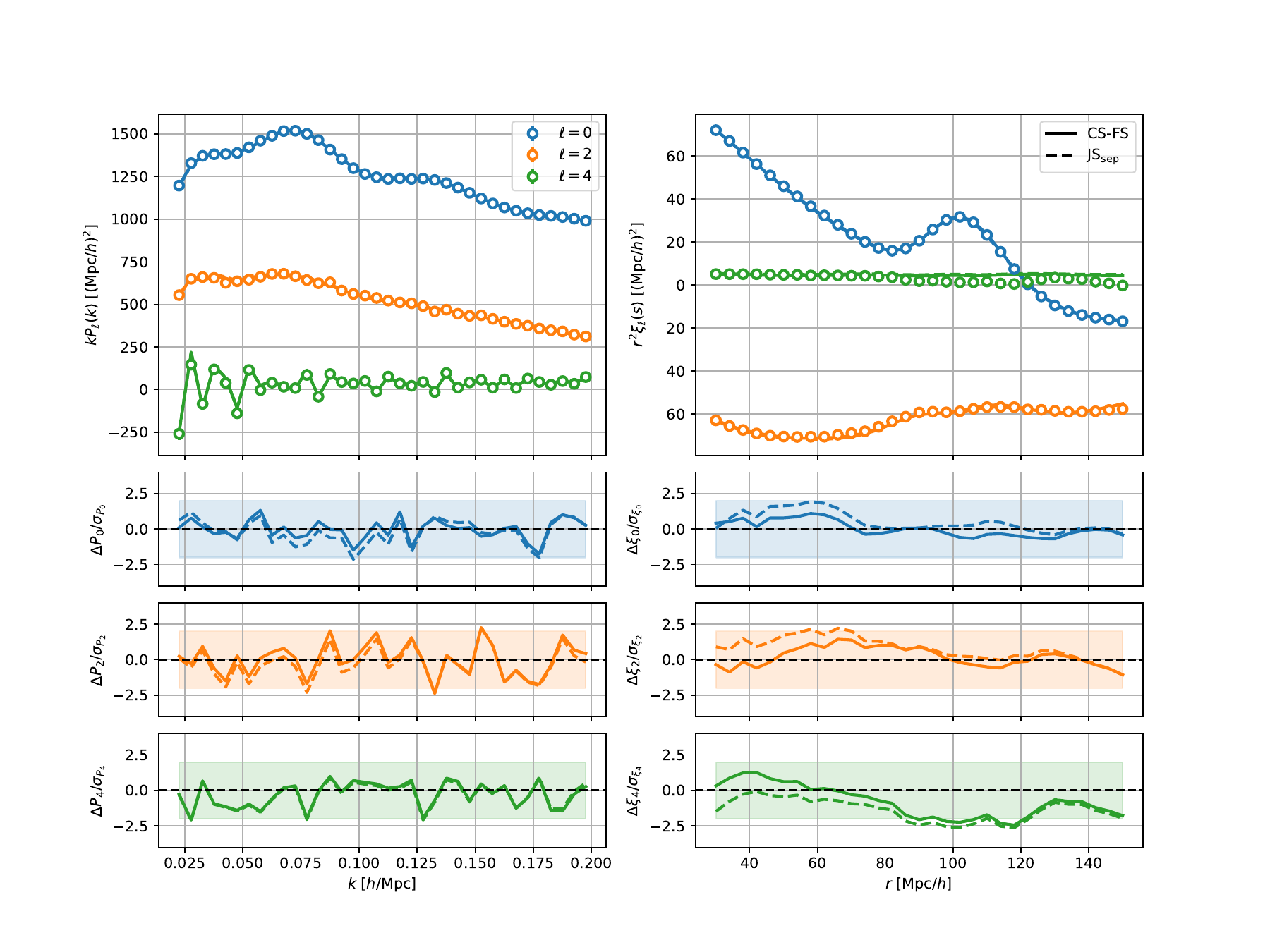}
    \caption{Fourier space (FS) power spectrum multipoles (left panel) and correlation function (CS) multipoles (right panel) with their best-fit models. We consider the monopole (blue), quadrupole (orange) and hexadecapole (green) in both spaces. 
    The three bottom panels show the normalized residuals and the shaded areas correspond to $2\sigma$. 
    The best-fit model of FS and CS are shown in solid lines while joint-space $\mathrm{JS}_\mathrm{sep}$ model is represented with a dashed-line.}
    \label{fig:bestfit_LRG_CB}
\end{figure*}

Figure \ref{fig:scale_rmin} presents the results of RSD measurements as a function of \(r_\mathrm{{min}}\) in CS. 
The first four rows show the best-fit values for \(\qpar\), \(\qper\), \(dm\) and \(d f = f/f_{\text{fid}}\), where \(f_{\text{fid}} = 0.838\) is the growth rate of the simulation, computed from their corresponding fiducial parameters (Table~\ref{tab:cosmo_parameters_desi_eboss}).
These best-fit values are compared against their expected values: \(\qpar^{\text{exp}}, \qper^{\text{exp}} = 1\), \(d f^{\text{exp}} = 1\) while $dm$ is expecting to be $0$. 
Both \(\qpar\) and \(\qper\) present residuals below 1 percent for all $r_\mathrm{min}$ values, while \(d f\) shows deviations below 3 per cent, except for the smallest $r_\mathrm{min}$ where the deviation slightly exceeds 3 per cent. 

The mid panel of Figure \ref{fig:scale_rmin} displays the reduced chi-squared, \({}_r\chi^2 \equiv \chi^2/(N_\mathrm{bins}-N_\mathrm{par})\). 
As expected, the \({}_r\chi^2\) values increase for decreasing \(r_\mathrm{{min}}\). 
Scales below \(30\,\rhMpc\) exhibit \({}_r\chi^2 > 3\), indicating a poor fit and suggesting that the model breaks down on such scales. 
As \(r_\mathrm{{min}}\) increases to approximately \(40\,\rhMpc\), the \({}_r\chi^2\) stabilizes around 1, reflecting improved agreement between the model and the data.

The last four rows of Figure~\ref{fig:scale_rmin} display 
the uncertainties \(\sigma_{\qpar}\), \(\sigma_{\qper}\), \(\sigma_{df}\) and \(\sigma_{dm}\) of the parameters of interest. 
Uncertainties naturally become larger with increasing \(r_\mathrm{{min}}\).
Consistent with past works, all values of \(\sigma_{\qpar}\) are approximately twice as large as those for \( \sigma_{\qper}\).

Our baseline choice for CS analysis aims at maximizing the information extracted from the correlation function while obtaining unbiased results on cosmological parameters and a good quality of fit. 
Based on the results just discussed, we set our baseline choice for our CS analysis to \(r_\mathrm{{min}} = 30\,\rhMpc\). 
{It is worth noting that we also tested the inclusion of stochastic terms, similar to those used in Fourier space, within the configuration space analysis. This addition helps regularize small scales beyond \(30\,\hmpc\), reducing the reduced \(\chi^2\) for \(r_{\mathrm{min}} = 25\,\hmpc\) to 1.6.}

\begin{figure*}[t]
    \centering
    \includegraphics[clip, trim=0.5cm 0.5cm 0 0 ,width=1\textwidth]{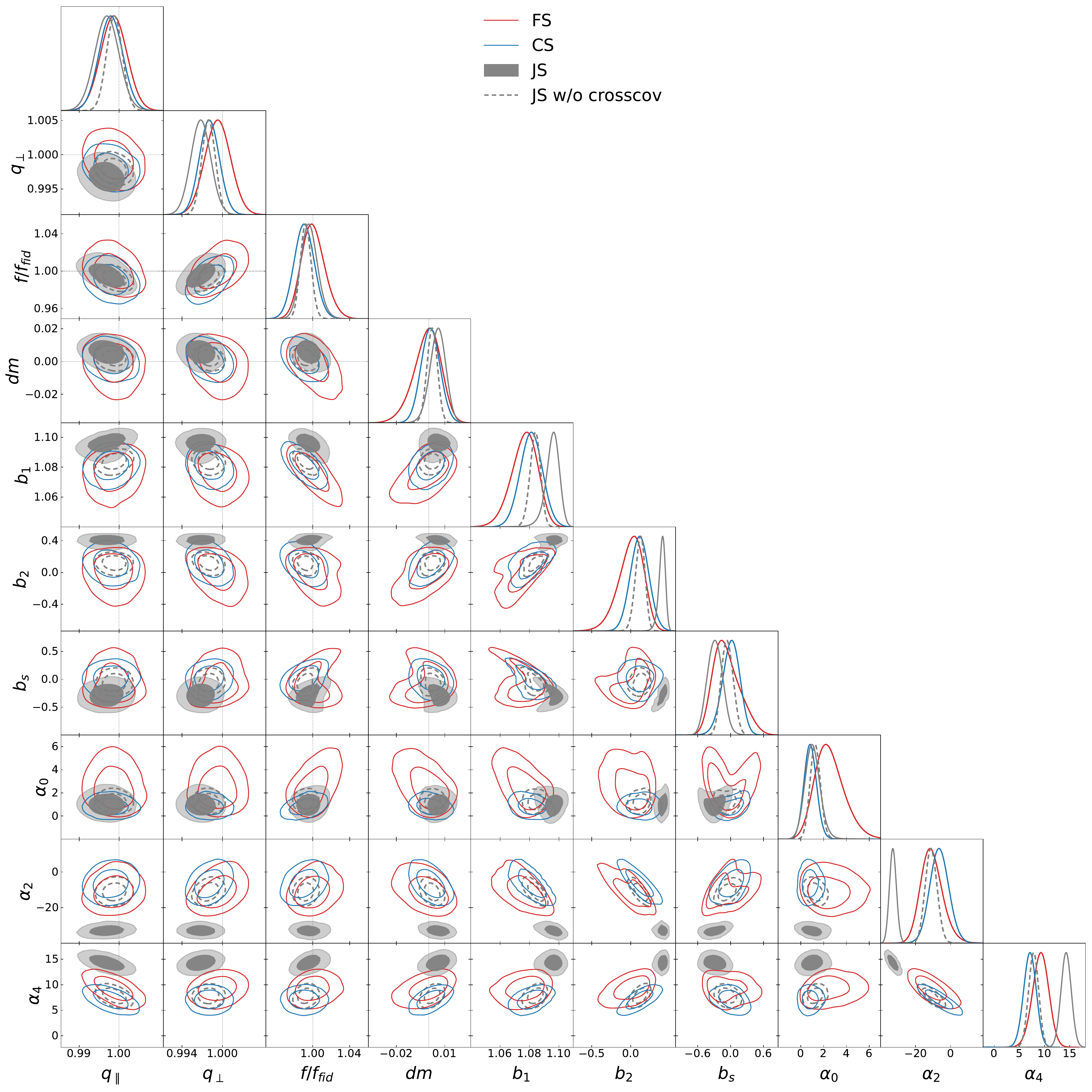}
    \caption{Likelihood posterior (\(1\sigma\)-\(2\sigma\)) for FS (red), CS (blue), and JS (grey) obtained from fits to the average of the 25 LRG CubicBox mocks. JS contours without the cross covariance (labelled "w/o crosscov") are displayed in dashed-grey. The dotted black lines represent the expected values. Results are shown for \(\kmax = 0.20 \, \khMpc\) and \(\rmin = 30 \, \rhMpc\).}
    \label{fig:mcmc_CS_FS}
\end{figure*}

Figure~\ref{fig:scale_kmax_total} presents the similar analysis as Figure~\ref{fig:scale_rmin} but for Fourier space (FS) and both joint space analyses (JS and JS$_\mathrm{sep}$). Here, we are interested in the impact of varying the maximum wavenumber \(\kmax\). 

Focusing first on FS results in Figure~\ref{fig:scale_kmax_total}  (red points), both \(\qpar\) and \(\qper\) show deviations below 1 per cent and \(d f\)  below 3 per cent for all \(\kmax\). 

The reduced chi-squared \({}_r\chi^2\) remains stable, slightly below 1.4, comparable to CS results at \(\rmin = 30 \rhMpc \). 
The uncertainties on \(\sigma_{\qper}\) and \(\sigma_{f}\) are notably larger than in CS, expect for \(\kmax \geq 0.22 \, \khMpc\) where they are comparable. 
Note that FS includes three additional parameters (shot-noise terms), 
which reduce its constraining power relative to the CS analysis. 
In the following, we restrict our analysis to \(\kmax = 0.20 \, \khMpc\), a choice that will be justified later when discussing the joint results. Figure \ref{fig:scale_kmax_total} also presents the joint-space results, which will be discussed later.

Focusing now in the JS analysis in Figure~\ref{fig:scale_kmax_total}  (grey circles), we observe small shifts on the best-fit values of all three parameters compared to FS or CS, though not significant. 
Regarding estimated uncertainties, JS yields uncertainties comparable to, or smaller than, those in CS, indicating some small gain in the combination with FS.  
However, \({}_r\chi^2\) values are significantly higher for JS, with values above 2.5 and increasing with $\kmax$, indicating a poor fit at this statistical precision level. 
As we will discuss next, the quality of the JS fit is degraded due to tensions in some of the nuisance parameters between CS and FS. For this reason, we perform a modified joint analysis where parameters \(b_2\), \(b_s\), \(\alpha_0\), \(\alpha_2\), and \(\alpha_4\) are allowed to vary independently in the power spectrum and correlation function models. 
This alternative approach is referred to as JS\(_\mathrm{sep}\). 
Figure \ref{fig:scale_kmax_total} presents the \(\mathrm{JS}_\mathrm{sep}\) results (grey squares), as a function of \(\kmax\). The overall performance of the fit is comparable to the standard JS analysis, though we note significant a reduction in \({}_r\chi^2\) to \(\sim 1.5\), comparable to CS and FS fits. 
{For \(\kmax > 0.20 \, \khMpc\), the \(\mathrm{JS}_\mathrm{sep}\) results for \(f\) and \(dm\) exhibit some discrepancies.}
Given those results, we set the baseline choice of $\kmax = 0.20\, \khMpc$ for FS, JS and JS$_\mathrm{sep}$.

Figure \ref{fig:bestfit_LRG_CB} presents the best-fit models with our baseline choice of separations ($30 < r < 150\,\hmpc$ and $0.02 < k < 0.20\,\ihmpc$) of our two-point multipoles ($\ell = 0,2,4$), along with their respective residuals in the lower panels. The differences between the models and the measurements remain within a \(2\sigma\) deviation, except for the hexadecapole in CS. We remind that neighboring separations in CS measurements are strongly correlated (Figure~\ref{fig:corr_matrix}) so these deviations are not significant.

Figure \ref{fig:mcmc_CS_FS} shows the 68 and 95 per cent contours of all parameter posteriors for FS, CS and JS. Posteriors in both FS and CS are in good agreement for all parameters, though CS constraints tend to be tighter (e.g. $b_2,b_s,\alpha_0$).  
For JS, contours for \(\qpar\) and \(\qper\) exhibit slightly larger biases compared to individual analyses, although they remain within \(2\sigma\). Surprisingly, the JS constraints on \(b_1\), \(b_2\), \(\alpha_2\) and \(\alpha_4\) are inconsistent with the FS and CS results. This might be due to the breaking of degeneracies in a multi-dimensional parameter space. An important caveat is that our JS fits use a sample covariance from 1000 mocks, though the data vector is larger. This covariance is thus noisier for JS compared to FS or CS (see Table~\ref{tab:m1m2}), which might shift contours. Given the large cross-covariance between FS and CS (as illustrated in Figure~\ref{fig:corr_matrix}), posteriors might be more sensitive to sample noise. To test this hypothesis, we computed JS posterior while neglecting the cross-covariance between FS and CS multipoles. These posteriors are shown in dashed lines in Figure \ref{fig:mcmc_CS_FS}, now in good agreement with the FS-CS contours, as expected when combining two independent likelihoods. 
{We also tested fits on the average of 1000 \package{EZmocks} (both DESI and eBOSS versions), using covariance matrices derived from the same set of mocks. This guarantees a correct covariance treatment in the fit, as it is directly derived from the same mocks used for the average. For both \package{EZmocks} sets, no significant bias is observed between CS, FS, and JS analyses. All three posteriors are fully consistent, in contrast to the bias seen in Figure \ref{fig:mcmc_CS_FS} for JS when using cubic box mocks.}
{We then compared the results of Figure  \ref{fig:mcmc_CS_FS} with the best-fit values obtained using \package{iMinuit}. For all analyses, the \package{iMinuit} best-fit points lie at the center of the contours, except for $b_s$ and $b_2$ which can be due to the presence of local mmaximum in the likelihood. This supports the bias observed for $b_1$, $\alpha_2$, and $\alpha_4$.}
{We also test the impact of the finite number of mocks used to compute the covariance by using 500 instead of 1000 mocks. No significant difference was seen, which discards the impact of the limited number of mocks.}

{These tests could indicate that the cross terms in a covariance matrix derived from 1000 \package{EZmocks} are not accurate enough when fitting N-body simulations.}
\begin{figure}[t]
    \centering
    \includegraphics[trim=0.7cm 0.5cm 0.5cm 0.5cm, width=0.4\textwidth]{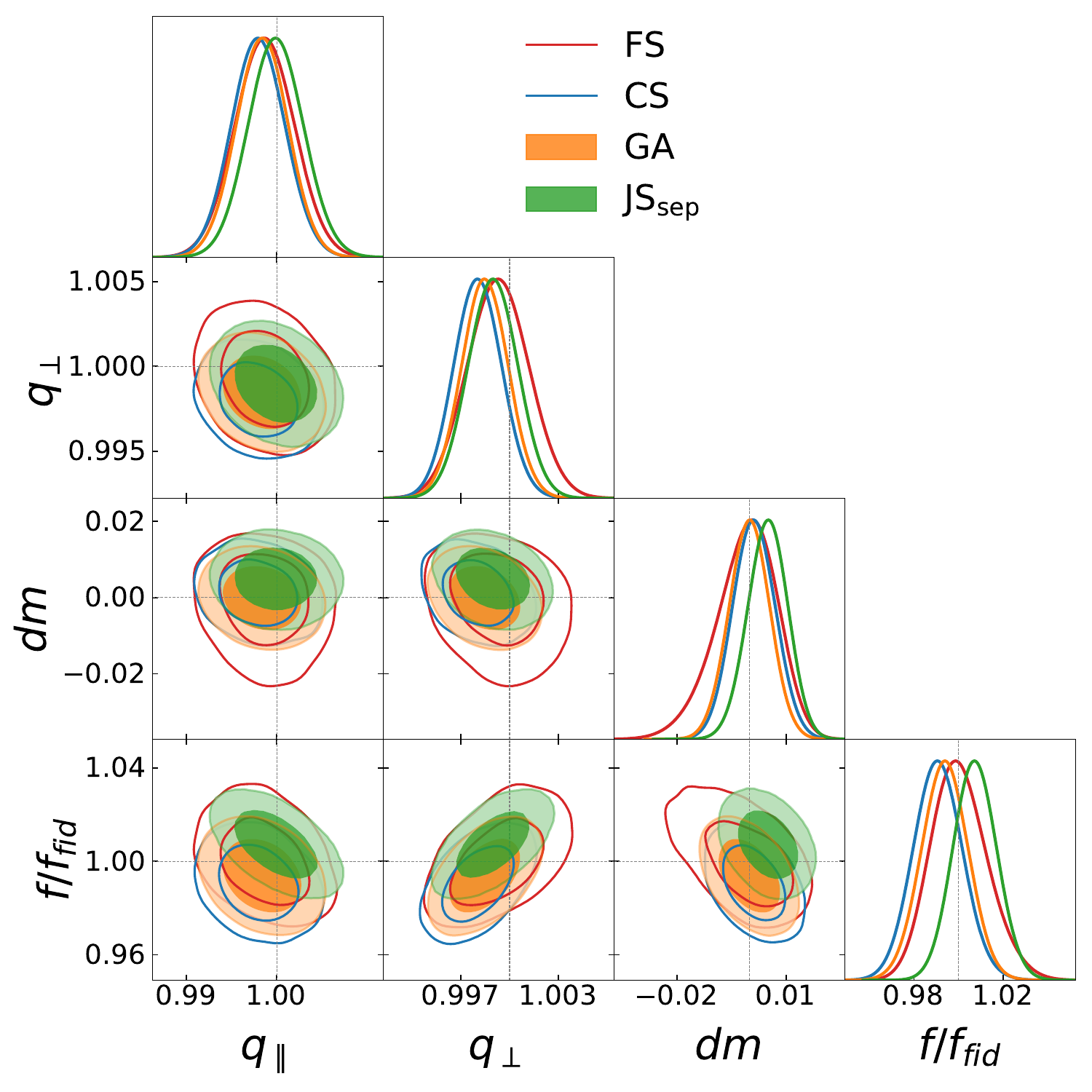}
    \caption{Contours of 68 an 95 per cent of the posterior distributions for FS (red), CS (blue), \(\mathrm{JS}_\mathrm{sep}\) (green), and GA (orange), derived from fits to the average clustering of 25 LRG cubic-box mocks from N-body simulations.}
    \label{fig:mcmc_GA}
\end{figure}

\begin{figure*}[t]
    \centering
    \includegraphics[clip, trim=0cm 0.5cm 0.2cm 0.3cm, width=0.85\textwidth]{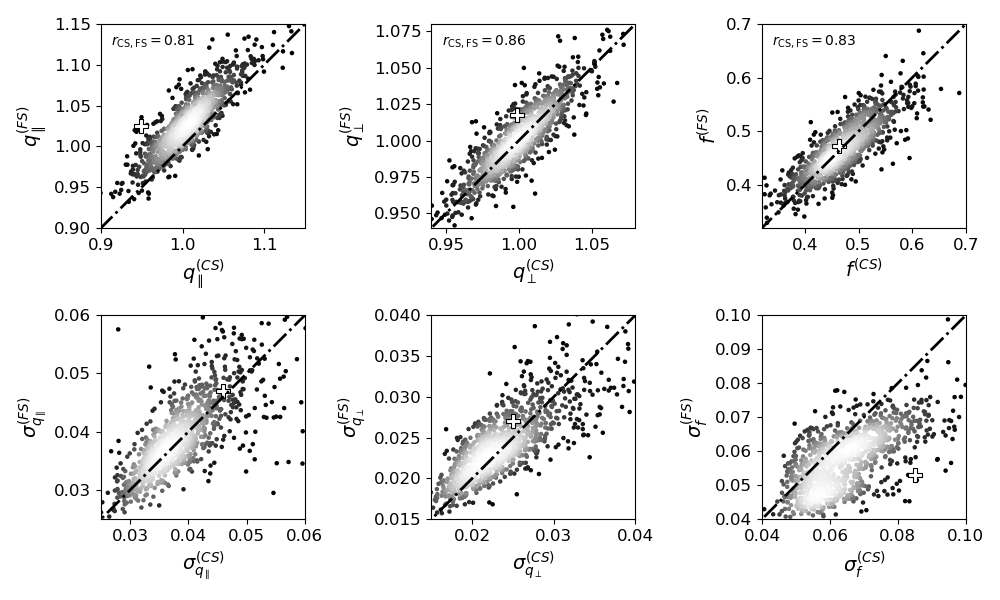}
\caption{Comparison of the RSD measurements in FS (y-axis) and CS (x-axis) for the individual 1000 \package{EZmocks} of the eBOSS LRG sample. The top row displays the best-fit values for \(\qpar\), \(\qper\), and \(f\), while the bottom row shows their estimated errors \(\sigma_{\qpar}\), \(\sigma_{\qper}\), and \(\sigma_{f}\). The density of points is represented by grayscale shading, with dashed lines indicating the line of perfect correlation where the compared parameters are equal. We display the Pearson correlation coefficients between CS and FS. The plus symbol represents the results from the real eBOSS data, as listed in Table \ref{tab:bestfit_eBOSS_LRG}.}
    \label{fig:ezmock_FS_CS}
\end{figure*}

Figure \ref{fig:mcmc_GA} presents posteriors for FS, CS (same as in Figure \ref{fig:mcmc_CS_FS}) as well as those from the \(\mathrm{JS}_\mathrm{sep}\) analysis and the Gaussian approximation method (GA, see section~\ref{sec:methodo:gaussian_approx}). 
All four contours are consistent with each other within \(1\sigma\). The GA contours tend to fall between the FS and CS contours, while the \(\mathrm{JS}_\mathrm{sep}\) do not show significant bias in the four parameters of interest to the separate analyses. 

Additionally, \(\mathrm{JS}_\mathrm{sep}\) does not show any notable improvement over GA when tested on the average of the 25 mocks, as both yield similar results.

These tests validate our baseline choices for the analyses and we turn to the study of a large set of mock catalogs.

\subsection{Fits to 1000 eBOSS LRG EZmocks}
\label{sec:results_mock:ezmocks}

We evaluated the statistical properties of our best-fit values and their uncertainties with fits to 1000 realizations of eBOSS LRG \package{EZmocks}, which reproduce realistic angular and redshift distributions (see section~\ref{sec:dataset}). 
These mocks allow us to assess how the results on individual realizations compare to their corresponding distributions, and check that our uncertainties are correctly estimated. 
We computed two-point multipoles for all realizations  and derived their corresponding sample covariance matrix (Eq.~\ref{eq:sample_covariance_matrix}). 

We used the frequentist approach to fit individual mocks (see section~\ref{sec:methodo}) as it is faster than the Bayesian approach.

Figure~\ref{fig:ezmock_FS_CS} compares the distributions of best-fit parameters in FS (y-axis) and CS (x-axis) for the 1000 individual eBOSS \package{EZmocks}. The top row displays the best-fit values for \(\qpar\), \(\qper\), and \(f\), 

while the bottom row shows their estimated errors \(\sigma_{\qpar}\), \(\sigma_{\qper}\), and \(\sigma_{f}\). The density of points is represented by grayscale shading, and the diagonal dashed line is the identity line. The distribution of the best-fit values is well correlated and mostly scattered around the identity line, except for \(\qpar\), which tends to have slightly higher values in FS than in CS.

{The discrepancy seen for $\qpar$ in FS may be due to non-Gaussian effects in the posteriors of parameters, which could be caused by the volume difference between the large effective volume of the covariance and the smaller volume of individual mocks. To mitigate this effect, we combine the 1000 \package{EZmocks} in groups of 5, making a total of 200 independent \package{EZmocks} with a volume 5 times larger. The discrepancy in FS is still present, which discards this hypothesis. We simply think that this bias is due to the approximate nature of \package{EZmocks} and that it is not guaranteed that the recovered cosmological parameters are aligned with the expected value, unlike \package{ABACUS} cubic boxes. We also performed a BAO fit on the average of the 1000 \package{EZmocks}, and both CS and FS recovered the expected value, while the full-shape fit on this average still shows the bias. This may indicate that when the fit is limited to higher scales as in the BAO case, the results are not biased, which confirms that the issue is related to the approximations in \package{EZmocks}.}

The uncertainties show good agreement overall for 
\(\qpar\) and \(\qper\), 
but for \(f\), CS uncertainties are larger than FS on average.
\begin{figure*}[t]
    \centering
    \includegraphics[clip, trim=0.5cm 0.5cm 0.2cm 0.3cm, width=0.85\textwidth]{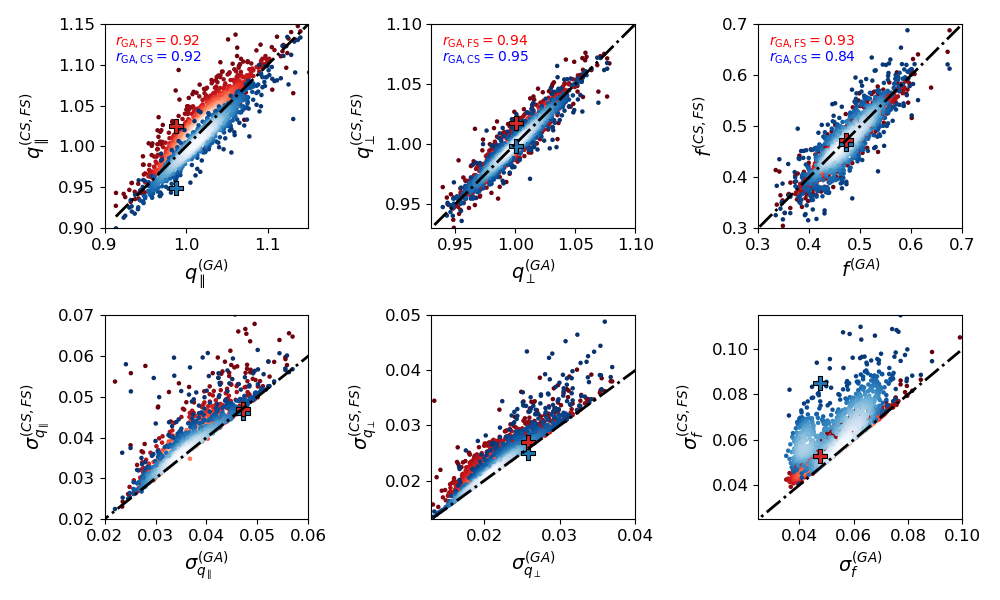}
        \caption{Comparison of the RSD measurements in FS (red) and CS (blue) with GA on the individual 1000 eBOSS EZmock. FS and CS results are shown on the y-axis, while GA results are on the x-axis. The top row displays the best-fit values for \(\qpar\), \(\qper\), and \(f\sigma_8\), while the bottom row shows their estimated errors \(\sigma_{\qpar}\), \(\sigma_{\qper}\), and \(\sigma_{f\sigma_8}\). The density of points is represented by the colorscale shading, and the dashed lines indicate the line of perfect correlation where the compared parameters are equal. We display the Pearson correlation coefficients between GA and FS, as well as GA and CS. The plus symbols represent the best-fit results from the real eBOSS data as listed in Table \ref{tab:bestfit_eBOSS_LRG}.}
    \label{fig:ezmock_GA}
\end{figure*}

\begin{figure*}[t]
    \centering
    \includegraphics[clip, trim=0.5cm 0.5cm 0.2cm 0.3cm, width=0.85\textwidth]{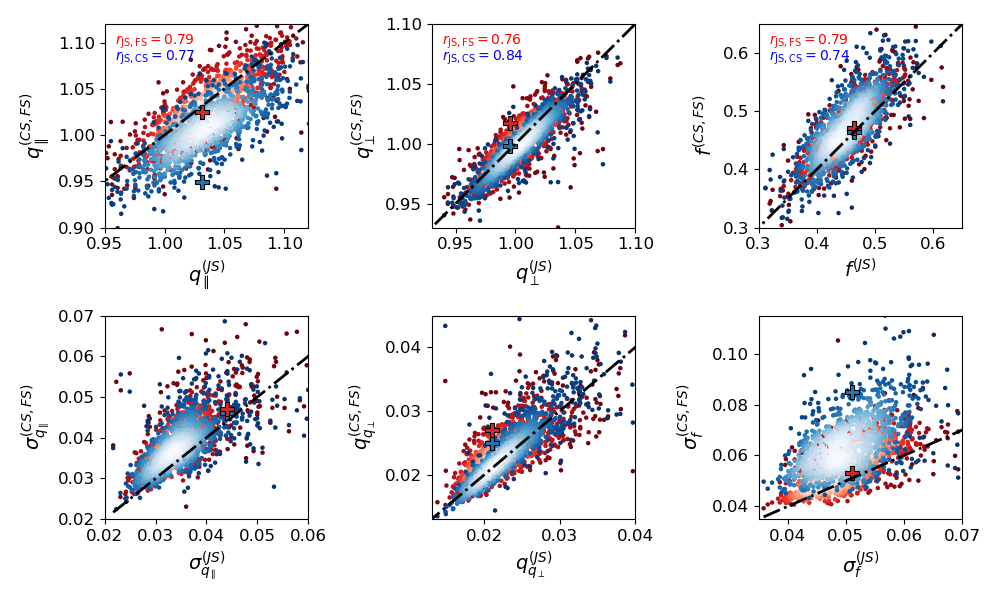}
\caption{Same as Figure \ref{fig:ezmock_GA} but for the \(\mathrm{JS}_\mathrm{sep}\) method. 
}
    \label{fig:ezmock_JS}
\end{figure*} 

\begin{table*}[t]
\centering
    \caption{Statistics on the fit of the 1000 eBOSS LRG EZmocks realizations.}
    {\tiny 
\begin{tabular}{p{0.4cm}p{0.5cm}p{0.8cm}p{0.5cm}p{0.5cm}p{0.5cm}p{0.5cm}p{0.7cm}p{0.5cm}p{0.5cm}p{0.5cm}p{0.5cm}p{0.7cm}p{0.7cm}p{0.5cm}p{0.5cm}p{0.5cm}p{0.7cm}}
\hline
\hline
&  & & \multicolumn{5}{c}{$ \qpar $} & \multicolumn{5}{c}{\( \qper \)} & \multicolumn{5}{c}{\( f \)} \\
\cmidrule(r){4-8} \cmidrule(lr){9-13} \cmidrule(l){14-18}
& \( N_{\text{good}} \) & 
\( \langle {}_r {\chi^2_{\text{min}}} \rangle \) &
\( \Delta {\qpar} \) & 
\( \sigma_{\qpar} \) & 
\( \langle \sigma_{\qpar} \rangle \) & 
\( \sigma_{Z_{\qpar}} \) & \( \langle Z_{\qpar} \rangle \) &   
\( \Delta {\qper}\) & 
\( \sigma_{\qper} \) & 
\( \langle  \sigma_{\qper} \rangle \) &
\( \sigma_{Z_{\qper}} \) & \( \langle  Z_{\qper} \rangle \) &  
\( \Delta f \) & 
\(  \sigma_{f} \) & 
\( \langle \sigma_f \rangle \)  & 
\(  \sigma_{Z_f} \) & \( \langle Z_f \rangle \) \\
& & & \( [10^{-2}] \) & \( [10^{-2}] \) & \( [10^{-2}] \) & & &  
      \( [10^{-2}] \) & \( [10^{-2}] \) & \( [10^{-2}] \) & & & 
      \( [10^{-2}] \) & \( [10^{-2}] \) & \( [10^{-2}]\)  & &   \\
\midrule
 CS                         & $1000$ & $0.72$ & 
 $0.88$ & $3.96$ & $3.88$ & 1.02 & -0.04 & 
 $0.04$ & $2.54$ & $2.40$ & 1.07 & -0.03 &
 $3.48$  & $13.7$  & $14.3$ & 0.89 & -0.08    \\
 FS                         & $999$  & $1.09$ & 
 $3.16$ & $4.21$ & $3.97$ & 1.06 & -0.08 & 
 $0.20$  & $2.51$ & $2.44$ & 1.00 & -0.03 & 
 $2.94$  & $11.9$ & $12.6$ & 0.90 & -0.07 \\
 GA                         & $999$  & $-$    & 
 $1.95$ & $4.00$  & $3.65$ & 1.19 & -0.02 & 
 $0.10$  & $2.37$ & $2.24$ & 1.12 & -0.03 & 
 $4.14$  & $11.1$ & $11.9$ & 1.10 & -0.08 \\
 $\mathrm{JS}_\mathrm{sep}$ & $1000$ & $0.92$ & 
 $3.73$ & $4.08$ & $3.54$ & 1.09 & -0.06 & 
 $0.65$ & $2.98$ & $2.34$ & 1.31 & -0.02 & 
 -0.98 & $11.6$ & $11.1$ & 0.92 & -0.03 \\
\end{tabular}
}
\label{tab:ezmock}
\tablefoot{ \(N_{\text{good}}\) is the number of valid realizations after removing undefined contours and extreme values and errors. We show the mean value of the best-fit reduced \({}_r {\chi^2_{\text{min}}}\). For each parameter $p$, we show the average bias \(\Delta p = \langle p_i - p_{\text{exp}} \rangle\), the standard deviation of best-fit values \(\sigma_{p} = \sqrt{\langle p_i^2 \rangle - \langle p_i \rangle^2}\), and the average of the per-mock estimated uncertainties \(\langle\sigma_{p_i}\rangle\). The pull is defined as \(Z_{p_i} \equiv (p_i - \langle p_i \rangle)/\sigma_{p_i}\) and we show values for the mean of the pull, \(\langle Z_{p_i} \rangle\)  and its standard deviation \(\sigma_{Z_{p_i}}\).}
\end{table*}

We computed consensus results using the Gaussian approximation method (GA) for each mock realization. 
First, we construct the $6\times 6$ parameter covariance matrix from the 1000 eBOSS \package{EZMocks} best-fit values of \(\qpar\), \(\qper\) and \(f\). 
Figure~\ref{fig:Ezmock_FS_CS_coefcorr} displays the resulting correlation matrix, where we can notice in particular that the correlations for \(\qpar\), \(\qper\) and \(f\sigma_8\) are 61, 51 and 81 percent,  respectively, between CS and FS. 
Second, we computed consensus parameters and their corresponding covariances for each mock realization, adjusting each time the off-diagonal blocks of the parameter covariance (see section~\ref{sec:methodo:gaussian_approx}).

Figure \ref{fig:ezmock_GA} shows the distributions of \(\qpar\), \(\qper\), \(f\), and their associated errors \(\sigma_{\qpar}\), \(\sigma_{\qper}\), \(\sigma_{f}\), as estimated with the GA method, compared to the same quantities obtained from CS and FS. The distribution of the best-fit values is well correlated and generally scattered around the identity line, except for \(\qpar\), where the scatter between GA and CS is slightly smaller than that between GA and FS. This suggests a higher weight of CS in the GA analysis compared to FS. Additionally, the errors estimated by GA are systematically smaller than those obtained in either CS or FS analyses. It is worth noting that the distribution of \(\sigma_{f}\) splits into two distinct clouds, as shown in Figure \ref{fig:ezmock_FS_CS}.

Figure \ref{fig:ezmock_JS} is analogous to Figure \ref{fig:ezmock_GA} but for the results from the  \(\mathrm{JS}_\mathrm{sep}\) method. Similar to the GA results, the best-fit values are well-correlated and closely cluster around the identity line. However, the scatter is noticeably more pronounced in \(\mathrm{JS}_\mathrm{sep}\) compared to GA 

While  \(\mathrm{JS}_\mathrm{sep}\) uncertainties are consistently smaller than or similar to those observed in CS and FS, they exhibit greater scatter relative to the GA results shown in Figure \ref{fig:ezmock_GA}. 
This scatter is due to the different nature of the fit, which now incorporates the cross-covariance between FS and CS at the level of the two-point functions. These noisy terms (due to a limited number of mocks) are propagated to the final infered parameters.
The similar or smaller uncertainties show that there is gain in combining FS and CS results at the two-point level, even though they are highly correlated. 

Table \ref{tab:ezmock} summarizes the statistical properties of $(\qpar, \qper, f)$ for  CS, FS, GA, and \(\mathrm{JS}_\mathrm{sep}\). For each parameter $p$, we show the average bias with respect to its expected value, $\Delta p = \langle p_i - p_{\text{exp}} \rangle$, the standard deviation of best-fit values $\sigma_{p}$, the mean estimated error $\langle \sigma_{p_i} \rangle$, the mean of the pull \(Z_{p_i} \equiv  \frac{p_i - \langle p_i \rangle}{\sigma_{p_i}}\), and its standard deviation. 
If the parameter uncertainties are correctly estimated and follow a Gaussian distribution, we expect $\sigma_{p} = \langle \sigma_{p_i} \rangle$, 
\(\langle Z_{p_i} \rangle = 0\) and \(\sigma_{Z_{p_i}} = 1\). 
The table also includes the mean value of the reduced chi-square, ${}_r\chi^2_{\text{min}}$. From this table, we observe: 

\begin{itemize}
    \item the number of valid realizations, \(N_{\text{good}}\), is nearly maximal for all analyses, indicating that most fits have successfully converged. The discarded cases in FS and GA are due to the removal of outliers beyond the \(5\sigma\) threshold. For each method, the average minimum reduced chi-square \( \langle {}_r {\chi^2_{\text{min}}} \rangle \) is close to 1, suggesting that the majority of the mocks are well-modeled, with \(\mathrm{JS}_\mathrm{sep}\) exhibiting the closest value from 1.

    \item the mean estimated error, \(\langle\sigma_{p_i}\rangle\), is generally consistent with the standard deviation of the best-fit values, \(\sigma_{p}\), across all parameters and analyses, indicating no significant issues with the uncertainty estimations. Notably, the mean estimated errors for \(\mathrm{JS}_\mathrm{sep}\) are smaller than those for CS, FS, and GA (except for \(\qper\)) which means that there is small gain in performing the joint analysis.

    \item the standard deviations of the best-fit values, \(\sigma_{p}\), are consistently smaller in GA compared to \(\mathrm{JS}_\mathrm{sep}\) for all three parameters. It is important to note that these standard deviations are scaled by the correction factor \(\sqrt{m_2}\) (section~\ref{sec:methodo:inference}) and these corrections are different between FS, CS and JS.

    \item the average bias for \(\qpar\) is larger in \(\mathrm{JS}_\mathrm{sep}\) and FS (more than 3 per cent) when compared to GA and CS. The bias in  \(\mathrm{JS}_\mathrm{sep}\) might be pulled by the large bias seen in FS.  
    The GA results lie between FS and CS as expected, for all parameters. 
    For the growth rate $f$,  \(\mathrm{JS}_\mathrm{sep}\) has the smallest bias among all methods.

    \item the mean of the pull \(\langle Z_{x_i} \rangle\) is consistent with zero for all methods. 
    For \(f\), the smallest value of the mean pull is achieved by \(\mathrm{JS}_\mathrm{sep}\).
    The standard deviations of the \(\qper\) and \(\qpar\) pulls are slightly larger than 1 in all cases, suggesting that errors are slightly underestimated, with the most significant underestimation for \(\qper\) in \(\mathrm{JS}_\mathrm{sep}\), of about 30 per cent. 
    For \(f\), the standard deviation of \(\mathrm{JS}_\mathrm{sep}\) is slightly smaller than 1, indicating a slight overestimation of the error. However, \(\mathrm{JS}_\mathrm{sep}\) provides results closest to 1 compared to the other methods.
    
     \item {The high value of \( \sigma_{Z_{\qper}} \) observed in $\mathrm{JS}_\mathrm{sep}$ may be due to the limited number of mocks used for the covariance. We recomputed the same statistics presented in Table \ref{tab:ezmock} using a covariance matrix derived from twice fewer mocks, i.e., 500 mocks. The standard deviation of the pull for the three parameters in $\mathrm{JS}_\mathrm{sep}$ increased to $\sim 4$, which may indicate the necessity of using a larger number of mocks for the covariance when performing JS analyses to obtain more reliable uncertainty estimates.}

\end{itemize}

\begin{figure*}[t]
    \centering
    {\includegraphics[clip, trim=2cm 1cm 2.5cm 2.5cm, width=0.8\textwidth]{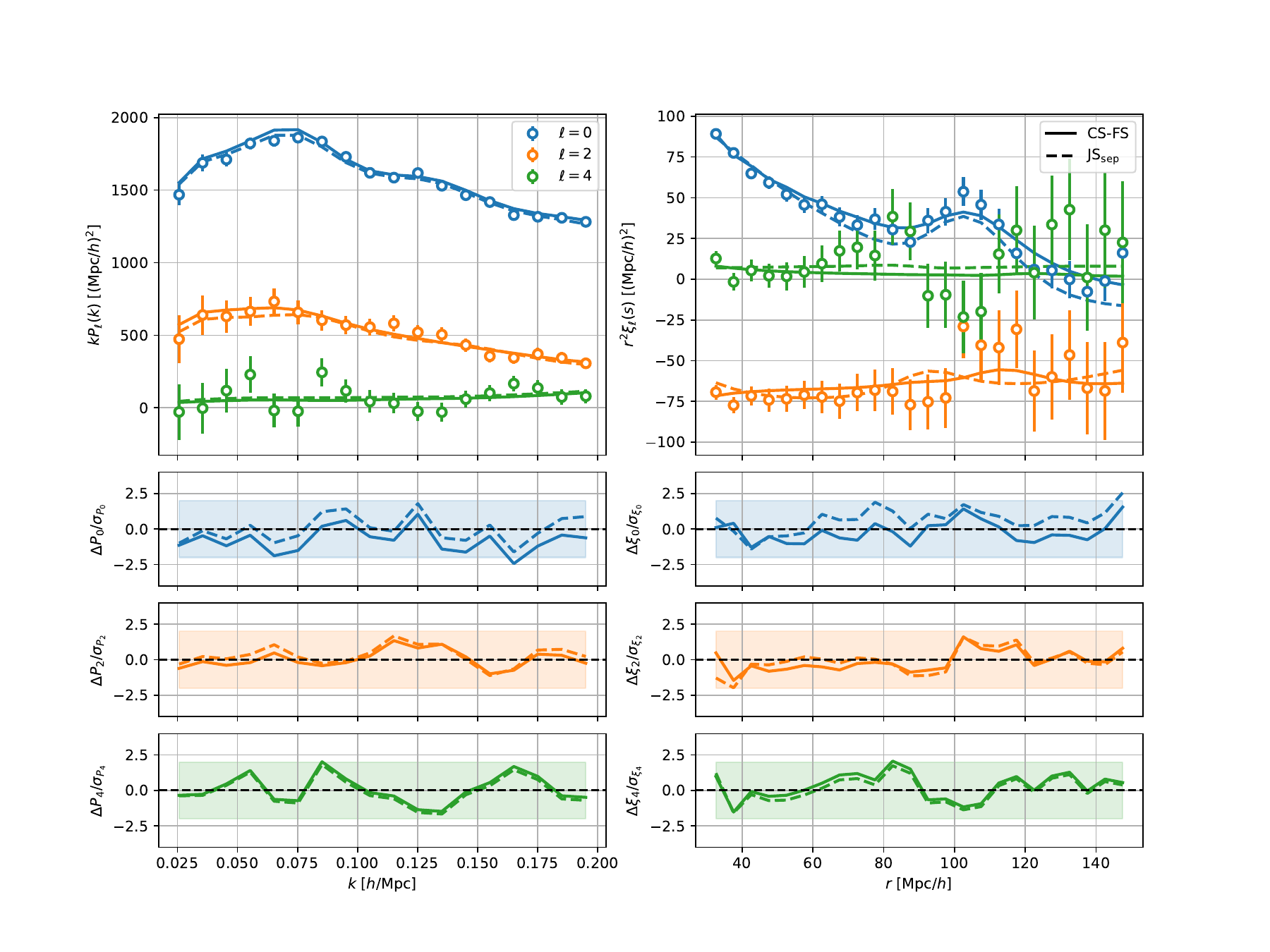}}
\caption{ Best-fit models for the monopole, quadrupole, and hexadecapole of BOSS+eBOSS LRG sample in Fourier-space (left panel, solid line) and configuration space (right panel, solid line). The bottom three sub-panels display the normalized residuals, 
with shaded areas indicating a \(2\sigma\) range. 
The best-fit model for joint space fit, \(\mathrm{JS}_\mathrm{sep}\), is shown as a dashed line.}
    \label{fig:bestfit_LRG_eBOSS}
\end{figure*}

\begin{figure}[t]
    \centering
  \includegraphics[width=0.4\textwidth]{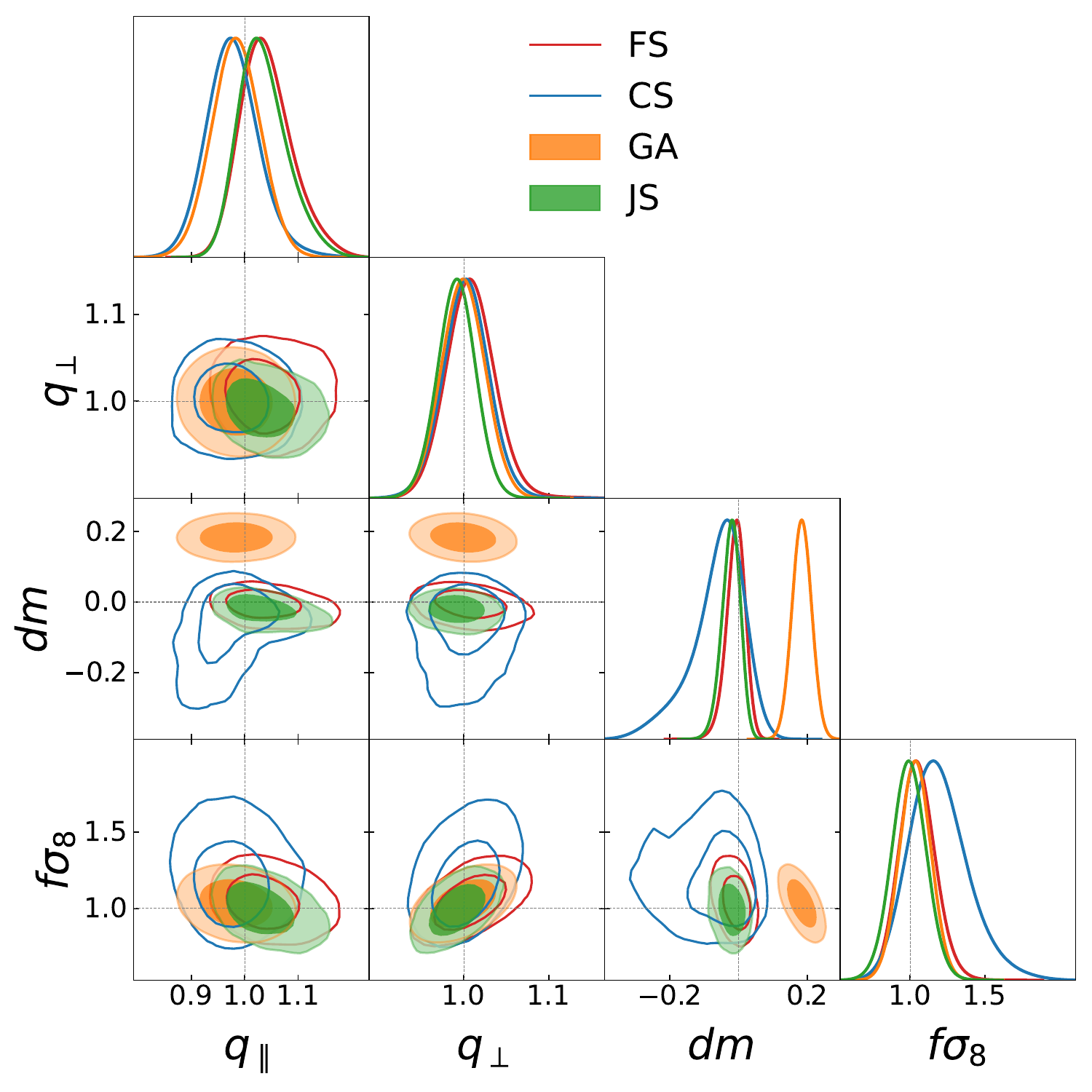}
\caption{   Likelihood posterior for FS (red), CS (blue), $\mathrm{JS}_\mathrm{sep}$ (green) and GA (orange) derived from fits on the eBOSS LRG data. }
    \label{fig:mcmc_eboss}
\end{figure}

\section{Application to the BOSS+eBOSS LRG sample}
\label{sec:results_eboss}

In this section, we apply our methodology to the BOSS+eBOSS LRG sample presented in section~\ref{sec:dataset:eBOSS}. We present results for Fourier space (FS), configuration space (CS), and our two consensus methods, the Gaussian approximation (GA) and joint space (\(\mathrm{JS}_\mathrm{sep}\)).

Figure \ref{fig:bestfit_LRG_eBOSS} displays the best-fit models for the power spectrum and correlation function multipoles for the separate analyses (FS and CS, shown as solid lines) and the joint analysis (\(\mathrm{JS}_\mathrm{sep}\), shown as dashed lines). The residuals are shown in the bottom panels, highlighting good agreement between \(\mathrm{JS}_\mathrm{sep}\) and CS/FS, which we attribute to differences in the inferred bias and counterterms.

Figure \ref{fig:mcmc_eboss} shows the contours of the posterior distribution ($1\sigma$ and $2\sigma$) for the FS, CS, GA, and \(\mathrm{JS}_\mathrm{sep}\) analyses, all of which are in good agreement. 
The posteriors are consistent with one another (within \(1\sigma\)), displaying comparable widths except for \(f\) in CS, which shows a higher value with a significantly larger width, though consistent with the behavior seen in eBOSS mocks (Figure \ref{fig:ezmock_FS_CS}), and the posterior of m for GA which is completly biased by more than \(2\sigma\))  
The \(\mathrm{JS}_\mathrm{sep}\) posterior is consistent with the other posteriors and comparable to GA, though we can see how these contours are not exactly Gaussian, which is one of the advantages of the method. 

Figure \ref{fig:mcmc_eboss} shows the contours of the posterior distribution ($1\sigma$ and $2\sigma$) for the FS, CS, GA, and $\mathrm{JS}_\mathrm{sep}$ analyses, all of which are in good agreement. 
The posteriors are consistent with one another (within $1\sigma$), displaying comparable widths except for $f$ in CS, which shows a higher value with a significantly larger uncertainty. This behavior is consistent with what we observe in the eBOSS mocks (Figure \ref{fig:ezmock_FS_CS}). The posterior of $m$ in GA, however, is significantly biased, deviating by more than $2\sigma$, illustrating the limitations of the GA method when the posteriors are non-Gaussian, as is the case for this parameter in CS. 
The $\mathrm{JS}_\mathrm{sep}$ posterior remains consistent with the other estimates and closely tracks GA, except for $m$, although its contours exhibit non-Gaussian features—highlighting one of the key strengths of the joint-space approach.

Table \ref{tab:bestfit_eBOSS_LRG} summarizes the results of the four different analyses. The reduced \(\chi^2\) values are 1.82, 1.10, and 1.34 for FS, CS, and \(\mathrm{JS}_\mathrm{sep}\), respectively. The high \(\chi^2\) value of FS represents a deviation greater than \(3\sigma\) compared to the distribution of \(\chi^2\) from \package{Ezmock} eBOSS. 
The best-fit values for the parameters \(\qpar\), \(\qper\), and \(f\sigma_8\) are all consistent within their respective \(1\sigma\) errors. 
For \(\qpar\) and \(\qper\), \(\mathrm{JS}_\mathrm{sep}\) provides the smallest uncertainties while yielding the highest and lowest best-fit values, respectively, comapred to the other methods. 
Additionally, \(\mathrm{JS}_\mathrm{sep}\) yields a slightly smaller value for \( f\sigma_8 \), similar to the one from CS.  
\begin{table*}[t]
\caption{Best-fit parameters from the eBOSS DR16 LRG sample using different methods. }
    \centering
    {\footnotesize
    \begin{tabular}{c  c c c c  c }
    \hline
    \hline
      Space &$\qpar $& $\qper$ & $dm$ & $f\sigma_8 $ & $\chi^2/dof$\\
    \hline
    FS   & $1.025 \pm 0.047$ & $1.017 \pm 0.027$ &  $0.031 \pm 0.060$ & $0.472 \pm 0.054$ & $75 / (54 - 13) = 1.82$    \\
    CS & $0.949 \pm 0.046$ & $0.998 \pm 0.025$ & $-0.180 \pm 0.167\,\,\,\,$ & $0.464 \pm 0.089$ & $69 / (72 - 10) = 1.10$   \\
    GA  & $0.988 \pm 0.047$  & $1.000 \pm 0.026$ &  $0.183 \pm 0.028$ & $0.476 \pm 0.048$ & $-$  \\
    \textbf{JS}$_\mathbf{sep}$ & $\mathbf{1.031 \pm 0.044}$ & $\mathbf{0.995 \pm 0.021}$ &$\mathbf{-0.041\pm 0.058}\,\,\,\, $ & $\mathbf{0.463 \pm 0.052}$ & $\mathbf{144 / (126 - 18) = 1.34}$ \\
    \hline
    FS eBOSS  &  $0.999 \pm 0.039$ & $1.003 \pm 0.030$ &-&    $0.454\pm 0.046$  & - \\
    CS eBOSS  & $1.013 \pm 0.034$ & $0.999 \pm 0.023$ &-  & $0.460 \pm 0.050$ & -  \\
    GA eBOSS & $1.009 \pm 0.034$  & $0.998 \pm 0.022$ &-&  $0.449\pm 0.044$ & - \\
    \end{tabular}}
    
    \label{tab:bestfit_eBOSS_LRG}
    
    \tablefoot{Fourier space (FS), configuration space (CS), Gaussian approximation (GA), and the joint-space analysis (\(\mathrm{JS}_\mathrm{sep}\)). These results are derived from the MCMC chains presented in Fig.~\ref{fig:mcmc_eboss} and correspond to the argmax values. FS eBOSS corresponds to the main results from \cite{gil-marinCompletedSDSSIVExtended2020} (Table 8, "FS $P_k$"), CS eBOSS represents the main results from \cite{bautistaCompletedSDSSIVExtended2021} (Table 14, "RSD $\xi_l$"), and GA eBOSS denotes the consensus results combining eBOSS analyses.}
\end{table*}
Still in Table \ref{tab:bestfit_eBOSS_LRG}, we compare our main \(\mathrm{JS}_\mathrm{sep}\) results with the official RSD results from \citet{gil-marinCompletedSDSSIVExtended2020} (FS eBOSS), \citet{bautistaCompletedSDSSIVExtended2021} (CS eBOSS), 
and the consensus between the two (GA eBOSS). 
The main difference between our analysis and official ones lies in the modeling approach. 
Previous analyses employed the TNS model \citep{taruya_baryon_2010}, which is based on an Eulerian description and incorporates a damping term for Fingers-of-God rather than counterterms. The scale ranges are the same for both CS analyses while for FS, we increased from $\kmax = 0.15$ to $0.20\,\ihmpc$. 

Our final result on this dataset using our \(\mathrm{JS}_\mathrm{sep}\) method yields:
\begin{equation}
  \left\lbrace \begin{array}{l}
\qpar = 1.031 \pm 0.044 \\
 \qper = 0.995 \pm 0.021 \\
f\sigma_8 = 0.463 \pm 0.052 
\end{array}\right.
\end{equation}
 Our results are compatible with all three analyses but show a slightly lower value for \(\qpar\), a higher value for \(\qper\), and for \(f\sigma_8\). These discrepancies can largely be attributed to differences in the clustering models, the scale ranges, the inclusion of cross-covariance terms in the JS analysis, and the inclusion of systematic errors (in the case of the official ones). 
 Uncertainties on \(\qpar\) are notably larger than in the official eBOSS results, likely due to the inclusion of additional counterterms and shot noise terms in our model, which introduce \(\mu^2\) and \(\mu^4\) dependencies (see Equation \ref{eq:model_pk_velo}).

\section{Conclusion}

This work presents a new RSD analysis of DR16 LRG sample (BOSS+eBOSS), where we extract cosmological parameters simultaneously from the configuration-space and Fourier-space two-point multipoles. We compared the joint-space (JS) approach to obtain consensus results with the commonly used Gaussian approximation (GA), which combines CS and FS results at the parameter level, assuming Gaussian posteriors. 

To validate our methodology, we first applied it to the average of 25 N-body LRG cubic-box mocks from the AbacusSummit simulation. We assessed parameter consistency between CS and FS and compared JS results with each individual space. Unexpected behavior was observed for parameters such as $b_2$, $\alpha_2$, and $\alpha_4$, with $1\sigma$–$2\sigma$ deviations from CS and FS, possibly due to noisy cross-terms in the covariance matrix. Allowing parameters $b_2$, $b_s$, and the counterterms to vary independently in JS (\(\mathrm{JS}_\mathrm{sep}\)) led to unbiased results with a reduced ${}_r\chi^2 = 1.5$. However, JS did not show significant improvements over CS, FS, or GA when tested on the mock average. Further tests were conducted on 1000 LRG EZmocks, which reproduce the DR16 LRG sample geometry. We found JS results to be consistently smaller than those of CS and FS but with greater scatter compared to GA.

After validation of our framework, we applied it to real data of the DR16 LRG multipoles. The JS method provided the tightest constraints on $\qpar$, $\qper$, and $f\sigma_8$, remaining consistent with the official 2020 results from  \citet{bautistaCompletedSDSSIVExtended2021} and  \citet{gil-marinCompletedSDSSIVExtended2020}. However, \(\mathrm{JS}_\mathrm{sep}\) yielded the highest value for \(f\sigma_8 = 0.463 \pm 0.052\), which we attribute to differences in modeling and the inclusion of cross-covariance terms between FS and CS.

We believe our framework is more robust than FS, CS or GA approaches and should be used in future surveys. Systematic effects in either space are averaged out at the multipole level, providing unbiased constraints on the growth rate and dilation parameters. Future work employing our methods shall give some attention to the precision of covariance matrices estimates, as the large noise might affect the final constraints.

\begin{acknowledgements}
The project leading to this publication has received funding from 
Excellence Initiative of Aix-Marseille University - A*MIDEX, 
a French ``Investissements d'Avenir'' program (AMX-20-CE-02 - DARKUNI).

Thanks to NERSC.

\end{acknowledgements}

\bibliographystyle{aa}
\bibliography{reference.bib}

\appendix 

\section{Covariance matrix}

\begin{figure}[t]
    \centering
    \includegraphics[clip, trim=0.3cm 1.5cm 0.5cm 2cm,width=\columnwidth]{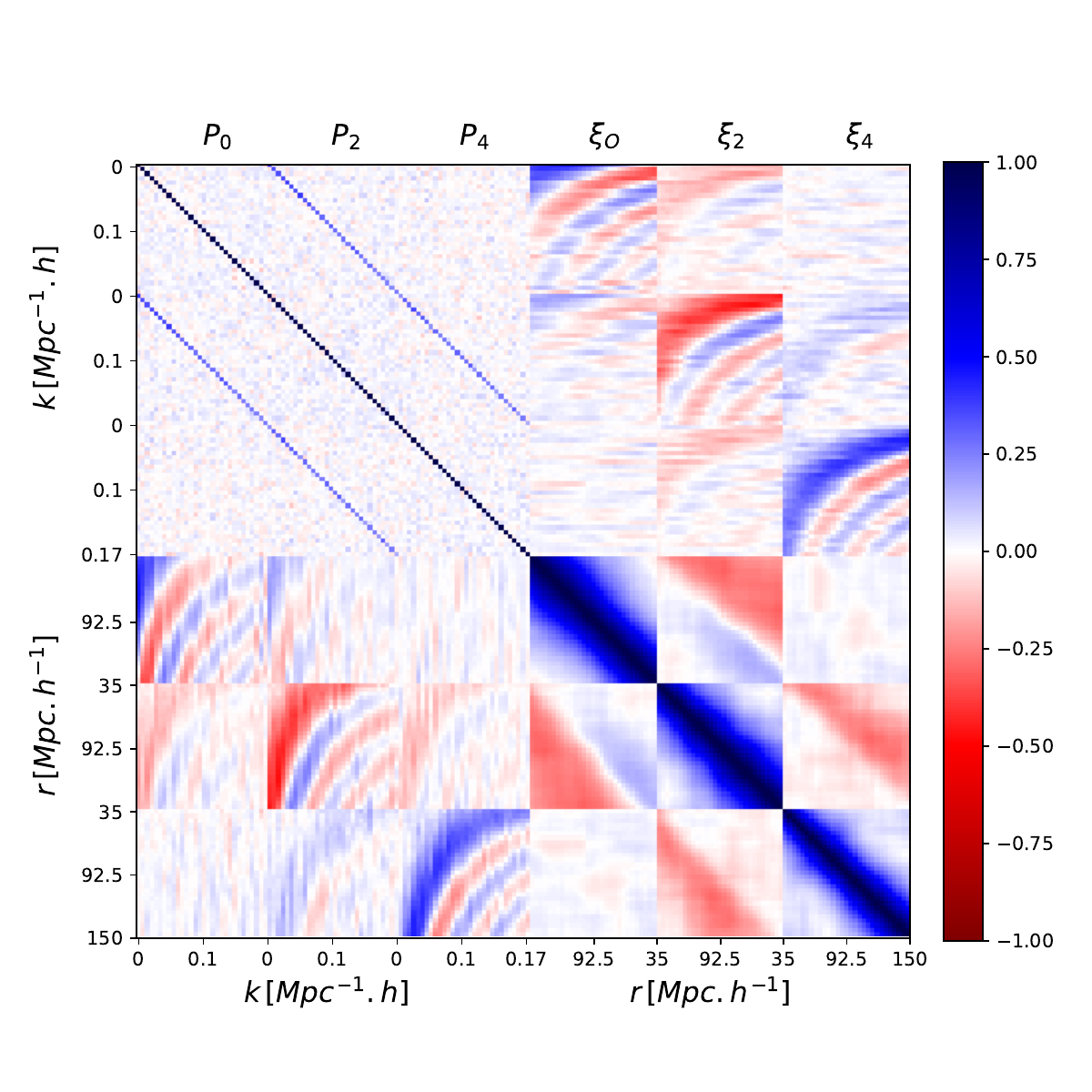}
    \caption{Normalised covariance matrix of the power spectrum multipoles \( P_{0,2,4} \) and the correlation function multipoles \( \xi_{0,2,4} \), measured from the 1000 cubic-box \package{EZmocks} of LRGs.}
    \label{fig:corr_matrix}
\end{figure}

Figure \ref{fig:corr_matrix} shows the correlation matrix \(\bm{R}\), defined as 
\begin{equation}
\bm{R} = \bm{\sigma}^{-1}\bm{\hat{C}} [\bm{\sigma}^{-1}]^T,
\end{equation}
where \(\bm{\sigma}\) is derived from the square root of the diagonal elements of the covariance matrix \(\bm{\hat{C}}\). 
Beyond the well-known correlations between multipoles in the same space,  the structure of the off-diagonal blocks reveals the correlation dynamics between the two distinct spaces. The multipole-multipole correlations between these spaces exhibit similar patterns. This arises due to the Fourier transform, which links a specific physical scale \(r\) with multiple spectral modes \(k\). At a given scale \(r\), there are notable correlations (and anti-correlations) with several \(k\) modes. This indicates that statistical information from a physical scale is distributed across multiple spectral modes rather than being confined to a single mode. These oscillations seen in the covariance can be derived analytically following \citet{grieb_gaussian_2016} and shown in Appendix C of \cite{dumerchat_baryon_2022}.

\begin{figure}[t]
    \centering
    \includegraphics[clip,trim=0.5cm 1.5cm 0.2cm 2cm,width=\columnwidth]{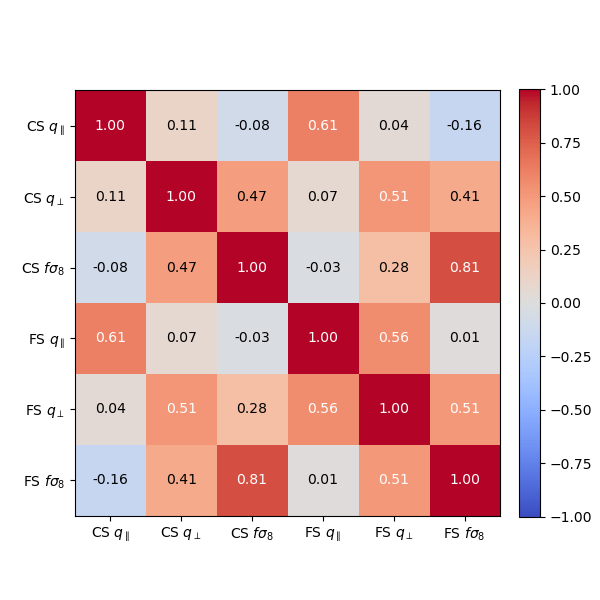}
\caption{Correlation coefficients between $\qper$, $\qpar$, and $f\sigma_8$ for CS and FS, derived from fits to the 1000 \package{EZmocks} of the eBOSS LRG sample.}
    \label{fig:Ezmock_FS_CS_coefcorr}
\end{figure}

Figure~\ref{fig:Ezmock_FS_CS_coefcorr} shows the correlation matrix of the parameters $\qpar,\qper$ and $\fsig$, obtained from 1000 fits to the eBOSS LRG EZmocks. This matrix is used to obtain the Gaussian approximation consensus results in section~\ref{sec:results_mock}.

\section{Clustering model}
\label{sec:model_app}

We use a Lagrangian perturbation theory (LPT) model to describe the observed clustering in both configuration and Fourier space. This model is based on the Lagrangian formalism developed in \citet{chen_consistent_2020,chen_redshift-space_2021}. In this section, we provide a brief overview of this formalism.

\subsection{Lagrangian Perturbation Theory}
\label{sec:model:lpt_app}

In the Lagrangian picture, cosmological structure formation is modeled by following the trajectories of fluid elements rather than studying the dynamics of density and velocity fields. The variable of interest in LPT is the displacement field \(\bm{\Psi} (\bm{q}, \eta)\), which corresponds to the difference between the Lagrangian coordinates \(\bm{q}\) (at some initial time \(\eta_0\)) and the Eulerian coordinates \(\mathbf{x}(\bm{q})\) of a fluid element at a given conformal time \(\eta\):  
\begin{equation}\label{eq_sec_2:displament_def_app}
\mathbf{x}(\bm{q}, \eta) = \bm{q} + \bm{\Psi} (\bm{q}, \eta),
\end{equation}
where \(\bm{\Psi}(\bm{q}, \eta_0) = 0\). Each element of the fluid is labeled by \(\bm{q}\), and \(\bm{\Psi} (\bm{q}, \eta)\) fully describes its evolution. The Eulerian density field \(\rho(\mathbf{x}, \eta)\) satisfies the continuity relation:
\begin{equation}\label{eq_sec_2:LPT_cont_eq_app}
\rho(\mathbf{x}, \eta) d^3x = \bar{\rho}(\eta) \, d^3q,
\end{equation}
where \(\bar{\rho}(\eta)\) is the mean density in comoving coordinates.

We can define the Jacobian of the transformation between Eulerian and Lagrangian coordinates as \(J = \det\left[\frac{\partial \mathbf{x}}{\partial \bm{q}}\right]\):
\begin{equation}\label{eq_sec_2:LPT_jacob_app}
d^3x = J(\bm{q}, \eta) \, d^3q.
\end{equation}

Using the definition of the density contrast \(\rho(\mathbf{x}, \eta) = \bar{\rho}(\eta) \left[1 + \delta(\mathbf{x}, \eta)\right]\) and Eqs.~\ref{eq_sec_2:LPT_cont_eq_app} and \ref{eq_sec_2:LPT_jacob_app}, we can relate the density contrast to the displacement field:
\begin{equation}
\delta(\mathbf{x}, \eta) = \frac{1}{J(\bm{q}, \eta)} - 1.
\end{equation}

This relation is equivalent to the following equation:
\begin{equation}\label{eq_sec_2:delta_k_LPT_1_app}
    1 + \delta(\mathbf{x}, \eta) = \int d^3q \, \delta_D\left[\mathbf{x} - \bm{q} - \Psi(\bm{q}, \eta)\right].
\end{equation}
Fourier transforming Eq.~\ref{eq_sec_2:delta_k_LPT_1_app} (in \(x\)-space) gives:
\begin{equation}
\label{eq_sec_2:delta_k_LPT_app}
    \delta(\bm{k}, \eta) = \int d^3q \, e^{i\bm{k} \cdot \bm{q}}\left(e^{i\bm{k} \cdot \Psi(\bm{q}, \eta)} - 1\right).
\end{equation}
The equation of motion for particle trajectories \(\mathbf{x}(\eta)\) is:
\begin{equation}
\ddot{\mathbf{x}} + \mathcal{H} \dot{\mathbf{x}} = -\nabla_\mathbf{x} \Phi(\mathbf{x}).
\end{equation}
From which we can infer the evolution equation of $\bm{\Psi}$ using Eq. \ref{eq_sec_2:displament_def_app}: 
\begin{equation}
\label{eq_sec_2:evo_psi_app}
\ddot{\bm{\Psi}} + \mathcal{H} \dot{\bm{\Psi}} = -\nabla_\mathbf{x} \Phi(\mathbf{x}),
\end{equation}
where overdots indicate derivatives w.r.t. conformal time $\eta$. This equation is solved order by order in a perturbative way such that:
\begin{equation}
\label{eq:psi_lpt_app}
    \bm{\Psi} = \bm{\Psi^{(1)}}+\bm{\Psi^{(2)}}+\bm{\Psi^{(3)}} + ...
\end{equation}
The first order solution of LPT $\bm{\Psi^{(1)}}$ is known a the Zel'dovich approximation \citep{zeldovich_gravitational_1970}:
\begin{equation}\label{eq_sec_2:Zeldovich_app}
    \bm{\nabla}_{\bm q} \bm\Psi^{(1)} = - D_{+}(\eta)\delta_0(\bm q)
\end{equation}
or equivalently: 
\begin{equation}\label{eq_sec_2:Zeldovich_2_app}
\bm{\Psi}(\bm{q}, \eta) = iD_+(\eta) \int \frac{d^3k}{(2\pi)^3} \frac{\bm{k}}{k^2} \delta_0(\bm{k}) e^{i\bm{k} \cdot \bm{q}}.
\end{equation}
where $ D_{+}(\eta)$ is the linear growth factor  and $\delta_0(\bm{q})$ is the initial density contrast. The Zel’dovich approximation uses the linear displacement field to model cosmic structure formation. One can find more details on the Zel'dovich approximation in \citet{white_zeldovich_2014}. Higher order solutions are expressed as integrals of higher powers of the linear density field \citep{matsubara_recursive_2015}.

Unlike in SPT, there is no simple recursive solution for order-by-order expression in LPT due to the fact that the displacement fields in LPT are not irrotational in general \citep{rampf_recursion_2012,matsubara_recursive_2015}.

\subsection{Effective field theory of large-scale structures}
\label{sec:model:eft_app}

In this work we use the Effective Field Theory of large-scales structures (EFTofLSS). The main insight is that the LSS has various hierarchical distance scales. In the large scale limit, only the large-scale degrees of freedom are required, while the effects of unknown short-scale physics can be systematically captured by effective operators. This decoupling framework forms the core of the EFTofLSS \citep{baumann_cosmological_2012,carrascoEffectiveFieldTheory2012,cabassSnowmassWhitePaper2022,ivanovEffectiveFieldTheory2022}.
The Lagrangian form of EFT involves smoothing Equation \ref{eq_sec_2:evo_psi_app} and incorporating small-scale physics through a series of counterterms. 

To achieve this, we use a filter \(W_R(\bm{q}, \bm{q'})\), which separates the system into long-wavelength (L) and short-wavelength (S) modes:
\begin{align}
&\bm{\Psi}_L(\bm{q}) = \int d^3q' W_R(\bm{q}, \bm{q'}) \bm{\Psi}(\bm{q'})\\
&\bm{\Psi}_S(\bm{q}, \bm{q'}) = \bm{\Psi}(\bm{q'}) - \bm{\Psi}_L(\bm{q})     
\end{align}
we can write the displacement as:
\begin{equation}\label{eq_sec_2:LPT_psi_app}
\bm{\Psi}(\bm{q}) = \bm{\Psi}^{(1)}_L(\bm{q}) + \bm{\Psi}^{(2)}_L(\bm{q}) + \bm{\Psi}^{(3)}_L(\bm{q}) + \cdots + \frac{1}{2} \alpha_1 \bm{\nabla} \delta_0 + \bm{S} + \cdots
\end{equation}
where the first three terms come from the perturbative treatment of the long-wavelength evolution and the last two terms parameterize the impact of the short-wavelength modes on the evolution where $\alpha_1$ is a counterterm coefficient and $\bm{S}$ is equivalent to a stochastic term

\subsection{Redshift-space distortions}
\label{sec:model:rsd_app}

Before writing the power spectrum we need to include bias scheme. The density of biased tracers can be modeled, assuming Lagrangian bias, by multiplying the $\delta_D$ in the above equation by the bias functional $F\left[\delta_0(\bm q), \bm\nabla^2 \delta_0(\bm q), ...\right]$ depending on linear density and its derivative. The bias functionnal is described by Eq. 4.4 in \citet{chen_consistent_2020}, more details about the modelling of biased tracers in the Lagrangian picture can be found in \citet{matsubara_nonlinear_2008}, \citet{mcdonald_clustering_2009-1} and \citet{vlah_gaussian_2016}. Equation \ref{eq_sec_2:delta_k_LPT_app} simply becomes for biased tracers:
\begin{equation}
\label{eq:biaseddelta_app}
\delta(\bm{k},\eta) = \int d^3q \, e^{i\bm{k} \cdot \bm{q}}\left(e^{i\bm{k} \cdot \Psi(\bm{q},\eta)} F(\bm{q}) - 1\right).
\end{equation}
The power spectrum $P(\vec k)$ can be expressed by the displacement field using Eq.(\ref{eq:biaseddelta_app}): 
\begin{equation}
\label{eq:pk_app}
    P(\bm k) = \int d^3\bm q\, e^{i\bm k \cdot \bm q}\left(\left< e^{i\bm k \cdot \bm \Delta}\right> - 1\right)
\end{equation}
where we have defined the pairwise Lagrangian displacement $\bm\Delta_i = \bm\Psi_i(\bm q_1)- \bm\Psi_i(\bm q_2)$. The cumulant theorem allows us to write the expectation value of the exponential in terms of the exponential of expectation values. The cumulant theorem is well defined in \citet{bernardeauLargeScaleStructureUniverse2002, matsubara_nonlinear_2008,matsubara_resumming_2008-1} 
and can be expressed as : 
\begin{equation}
    \left<e^{-iX}\right>  = \exp\left[\sum_{N=1}^\infty \frac{(-i)^N}{N!} \left<X^N\right>_c\right]
\end{equation}
where $\left<X^N\right>_c$ is the cumulant of a random variable $X$. The bracketed average in Eq. (\ref{eq:pk_app}) can be expressed as : 
\begin{equation}
    \ln \left< e^{i\bm k \cdot \bm \Delta}\right>  = - \frac{1}{2}k_ik_jA_{ij} - \frac{i}{6} k_ik_jk_k W_{ijk} + ...
\end{equation}
where $A_{ij} = \left<\Delta_i\Delta_j \right>_c$ and $W_{ijk} = \left<\Delta_i\Delta_j\Delta_k \right>_c$ are the cumulants of the pairwise displacements. More details of those terms can be found on \citet{vlahLagrangianEffectiveField2015,chen_redshift-space_2021}.

The power spectrum in equation \ref{eq:pk_app} is valid at all times, not only on the linear regime. From \citet{taylor_nonlinear_1996} we can express the power spectrum in the Zel'dovich approximation. In the Zel'dovich approximation, the displacement field is assumed to be Gaussian at all times. The expectation value of the exponential in Equation \ref{eq:pk_app}
\begin{equation}
    \left< e^{i\bm k \cdot \bm \Delta}\right> = \exp(-k_ik_j\left[ \Psi_{ij}(0) -  \Psi_{ij}(\bm{q})\right])
\end{equation}
with $\bm{q} = \bm{q_1}-\bm{q_2}$ and:
\begin{equation}
    \Psi_{ij}(\bm{q})=\left\langle \Psi_i(\mathbf{q}_1)\Psi_j(\mathbf{q}_2) \right\rangle
\end{equation}
Using the Zel'dovich solution Eq.\ref{eq_sec_2:Zeldovich_2_app} we can relate the $\Psi_{ij}(\bm{q})$ to the linear power spectrum \citep{taylor_nonlinear_1996}:
\begin{equation}
  \Psi_{ij}(\bm{q})  = \frac{1}{(2\pi)^3} \int d^3k \, \frac{k_i k_j}{k^4} P_{lin}(k) e^{-i\bm{k} \cdot \bm{q}}
\end{equation}
This results to the Zel'dovich power spectrum: 
\begin{equation}
\label{eq_sec_2:pk_ZA_app}
    P(\bm k) = \int d^3\bm q\, e^{i\bm k \cdot \bm q}\times\left[e^{\left(-k_ik_j\left[ \Psi_{ij}(\bm{0}) -  \Psi_{ij}(\bm{q})\right] \right)} - 1\right]
\end{equation}
This equation describes the nonlinear transformation of the initial power spectrum into the evolved power spectrum.

Finally, in the Lagrangian formalism, the RSD can be accounted by boosting displacements along the line-of-sight (LOS) direction $\hat{n}$ by the velocities of galaxies : 

\begin{equation}
    \bm\Psi \to \bm\Psi^s = \bm\Psi + \frac{\hat{n}(\bm v \cdot \hat{n})}{\mathcal{H}}
\end{equation}
Where the superscript s is referring to vectors boosted into redshift space and $\bm v$ is the peculiar velocity of galaxy.  

The final expression of the theoretical model for the redshift-space galaxy power spectrum, accounting for bias in LPT and EFTofLSS is giving by Eq. \ref{eq:model_pk_velo}. 

\end{document}